# Towards a more efficient approach for the satisfiability of two-variable logic

TING-WEI LIN, CHIA-HSUAN LU, and TONY TAN, National Taiwan University, Taiwan

We revisit the satisfiability problem for two-variable logic, denoted by $\mathrm{SAT}(\mathsf{FO}^2)$, which is known to be NEXP-complete. The upper bound is usually derived from its well known *Exponential Size Model* (ESM) property. Whether it can be determinized efficiently is still an open question.

In this paper we present a different approach by reducing it to a novel graph-theoretic problem that we call *Conditional Independent Set* (CIS). We show that CIS is NP-complete and present two simple algorithms for it with run time $O(\delta_0^n)$ and $O(\delta_1^n)$, where $\delta_0 = 1.4423 \approx \sqrt[3]{3}$ and $\delta_1 = 1.6181 \approx (\sqrt{5}+1)/2$ and $n$ is the number of vertices in the graph. We also show that unless the *Strong Exponential Time Hypothesis* (SETH) fails, there is no algorithm for CIS with run time $\delta_2^n$, where $\delta_2 = \sqrt{2} \approx 1.4142$.

We then show that without the equality predicate $\mathrm{SAT}(\mathsf{FO}^2)$ is in fact equivalent to CIS in succinct representation. This yields two algorithms for $\mathrm{SAT}(\mathsf{FO}^2)$ without the equality predicate with run time $O(\delta_0^{(2^n)})$ and $O(\delta_1^{(2^n)})$, where $n$ is the number of predicates in the input formula. To the best of our knowledge, these are the first exact algorithms for an NEXP-complete decidable logic with run time significantly lower than $O(2^{(2^n)})$. We also identify a few lower complexity fragments of $\mathsf{FO}^2$ which correspond to the tractable fragments of CIS. Similar to CIS, unless SETH fails, there is no algorithm for $\mathrm{SAT}(\mathsf{FO}^2)$ with run time $O(\delta_2^{(2^n)})$.

For the fragment with the equality predicate, we present a linear time many-one reduction to the fragment without the equality predicate. The reduction yields *equi-satisfiable* formulas and incurs a small constant blow-up in the number of predicates. Finally, we also perform some small experiments which show that our approach is indeed more promising than the existing method (based on the ESM property). The experiments also show that although theoretically it has the worse run time, the second algorithm in general performs better than the first one.

## 1 INTRODUCTION

Two-variable logic ($\mathsf{FO}^2$) is one of the well known fragments of first-order logic that comes with decidable satisfiability problem, henceforth, denoted by $\mathrm{SAT}(\mathsf{FO}^2)$. The exact complexity is NEXP-complete [9, 12, 25, 29, 40] and the upper bound is usually derived from its well known *Exponential Size Model* (ESM) property which states that every satisfiable formula has a model with size at most exponential in the length of the input formula [12]. Thus, to decide whether a formula is satisfiable,

Authors' address: Ting-Wei Lin, r08922089@csie.ntu.edu.tw; Chia-Hsuan Lu, r09922064@csie.ntu.edu.tw; Tony Tan, tonytan@csie.ntu.edu.tw, National Taiwan University, Taipei, Taiwan, 10617.

it suffices to non-deterministically "guess" a model with exponential size and verify that it is indeed a model of the formula. This is the only known algorithm for $\mathrm{SAT}(\mathsf{FO}^2)$ and it is open whether it can be efficiently determinized. Enumerating all possible structures would be naïve and practically infeasible.[1]

In [19] a better determinization is proposed where $\mathsf{FO}^2$ formulas are encoded as Boolean formulas, which can then be fed into a SAT solver for satisfiability testing. This approach stems from the observation that checking whether an FO formula has a model with a certain fixed size can be reduced to Boolean SAT, since the number of all possible ground facts becomes fixed as well. However, adopting this technique for $\mathrm{SAT}(\mathsf{FO}^2)$ has some major drawbacks.

First, the upper bound provided by the ESM property is not tight. While the bound *is* tight when the equality predicate is absent, it is no longer the case when the equality predicate is present. One can easily write an $\mathsf{FO}^2$ formula that is satisfiable only by models of certain sizes, say, $2^n$, while the upper bound provided by ESM property is of the form $3m2^n$, for some $m$ which depends on the input formula. In such cases the $\mathsf{FO}^2$ solver still has to determine the correct size, and it does so by testing various sizes from 1 up to the maximum number possible. This can lead to poor performance especially when the formula is unsatisfiable or when the formula has only "big" models. In almost all our experiments, when the models have size more than 12, the solver in [19, 24] does not terminate (within 24 hours).

Second, the conversion to Boolean formulas produces a lot of long clauses. For example, a formula of the form $\forall x \exists y\ \beta$ is encoded as Boolean formula of the form $\bigwedge_{i=1}^{n} \bigvee_{j=1}^{n} \ell_{i,j}$, for some $n$ which can be exponential. Thus, there will be exponentially many clauses, each with exponentially many literals. Since most SAT solvers find such formulas challenging, this may lead to poor performance. In fact, typical benchmarks in SAT competition contain relatively few long clauses [1].

*Our contribution.* In this paper we propose an entirely different approach for $\mathrm{SAT}(\mathsf{FO}^2)$ by reducing it to a novel graph-theoretic problem that we call *Conditional Independent Set* (CIS). Briefly, an instance of CIS is a tuple of graphs $(G_0, G_1, \dots, G_m)$. All of them are over the same finite set of vertices where $G_0$ is undirected graph and the others $G_1, \dots, G_m$ are all directed graphs. The task is to decide if there is a non-empty independent set (not necessarily maximal) $\Gamma$ in $G_0$ such that every vertex in $\Gamma$ has an outgoing edge in $G_i$ to another vertex in $\Gamma$, for every $1 \leqslant i \leqslant m$.

We prove that CIS is NP-complete and present two simple algorithms for it with run time $O(\delta_1^n)$ and $O(\delta_2^n)$, where $\delta_0 = 1.4423 \approx \sqrt[3]{3}$ and $\delta_1 = 1.6181 \approx (\sqrt{5}+1)/2$ and $n$ is the number of vertices in the graph. Note that since CIS is NP-complete, it is unlikely that it has a polynomial time deterministic/randomized algorithm. In fact, we show that unless the *Strong Exponential Time Hypothesis* (SETH) [5, 17, 18] fails, there is no algorithm for CIS with run time $O(\delta_2^n)$, where $\delta_2 = \sqrt{2} \approx 1.4142$. It is also worth mentioning that although the second algorithm has theoretically worse run time, our experiment shows that in general it performs better than the first one.

We then show that without the equality predicate $\mathrm{SAT}(\mathsf{FO}^2)$ is equivalent to CIS in succinct representation in the sense that an $\mathsf{FO}^2$ formula with $n$ unary predicates can be reduced to an instance of CIS with at most $2^n$ vertices. The converse is also true that an instance of CIS with $n$ vertices can be encoded as an $\mathsf{FO}^2$ formula with $\log n$ unary predicates.

By applying the algorithms for CIS, we obtain the same type of algorithms for $\mathrm{SAT}(\mathsf{FO}^2)$ without the equality predicate with run time $O(\delta_0^{(2^n)})$ and $O(\delta_1^{(2^n)})$, respectively, where $n$ is the number of unary predicates in the input formula. To the best of our knowledge, these are the first exact algorithms for an NEXP-complete decidable logic with run time significantly lower than $O(2^{(2^n)})$.

[1] To illustrate this point, it will take almost 60 years for a 10 GHz computer to complete a program with run time $2^{2^n}$ when $n$ is only as small as 6.

It should be noted that even when the equality predicate is absent, SAT($\mathsf{FO}^2$) is already NEXP-hard [9, 25]. Thus, it is also unlikely that it has an exponential time deterministic/randomized algorithm. Similar to CIS, unless SETH fails, there is no algorithm for SAT($\mathsf{FO}^2$) with run time $O(\delta_2^{(2^n)})$.

At this point, it is worth comparing the run time of our approaches with the one obtained via the ESM property as suggested in [19, 24]. Let $\Phi$ be input formula with $k_1$ unary predicates and $k_2$ binary predicates The ESM property gives us domain size $3m2^n$, where $n = k_1 + k_2$ and $m$ is an integer that depends on the formula. This means that the constructed Boolean formula contains $k_1 \cdot 3m2^n + k_2 \cdot 9m^2 2^{2n}$ Boolean variables, which yields worst case run time $2^{(k_1 \cdot 3m2^n + k_2 \cdot 9m^2 2^{2n})}$. This is a a much higher run time than ours which is of the form $\delta^{2^n}$, for various $\delta < 1.7$.

For the fragment with the equality predicate, we present a linear time many-one reduction to the fragment without the equality predicate. The reduction yields *equi-satisfiable* formulas and incurs a small constant blow-up in the number of unary predicates. Note that due to the ESM property, with the equality predicate SAT($\mathsf{FO}^2$) is in NEXP. Since SAT($\mathsf{FO}^2$) is already NEXP-hard when the equality predicate is absent, it is implicit that there *exists* such a polynomial time many-one reduction. However, there is no known explicit reduction so far, and as far as we know, our reduction is the first one.

Finally, we perform small experiments comparing the performance of our algorithms with the one in [19, 24] and with Z3 solver [2]. In general our algorithms works better than the existing ones and the second algorithm always performs better than the first one, although the first has a better theoretical upper bound.

It is also worth noting that recasting SAT($\mathsf{FO}^2$) as a graph theoretic problem not only gives us more efficient and practical algorithms, but also fresh ideas on designing interesting benchmarks. These include *random* $\mathsf{FO}^2$ formulas, which are formulas obtained by first generating random graph systems (in the sense of the Erdös-Rényi model) and then constructing the corresponding $\mathsf{FO}^2$ formulas.

*Related works.* Scott [40] was the first to prove the decidability of SAT($\mathsf{FO}^2$) by reducing it to the so called Gödel class formula, though his proof only works for the fragment without the equality predicate. The decidability of the general SAT($\mathsf{FO}^2$) was first proved by Mortimer [29] by showing that every satisfiable $\mathsf{FO}^2$ formula has double-exponential size model. The bound was later improved to single exponential by Grädel, Kolaitis and Vardi [12], which immediately implies that SAT($\mathsf{FO}^2$) is in NEXP. Matching lower bound was established by Fürer [9], based on the work of Lewis [25]. In fact, the lower bound already holds for the fragment of formulas with prefix $\forall\forall \wedge \forall\exists$ using only unary predicates and without the equality predicate.

De Neville and Pratt-Hartman [8] proposed a resolution based algorithm for SAT($\mathsf{FO}^2$), but their algorithm does not come with any guaranteed time complexity. From the work of Kieronski, Otto and Pratt-Hartmann [22, 34, 36], satisfiability of extensions of $\mathsf{FO}^2$ with counting quantifier and one equivalence relation remains in NEXP. However, these algorithms are all non-deterministic, and it is not immediately clear to what extent they can be implemented efficiently.

There has been considerable research effort to establish efficient (deterministic/randomized) algorithms for Boolean $k$-SAT [14, 27, 32, 33]. For $k = 3$, well known algorithms such as by Monien and Speckenmeyer [27], Rodosek [38], Schöning [39] and PPZ/PPSZ/biased PPSZ [14, 32, 33], just to name a few, come with run time such as $O(1.619^n)$, $O(1.476^n)$, $O((4/3)^n)$, $O(1.364^n)$ and some tiny improvement on them. As far as we know, there is no similar line of work on NEXP-complete decidable logic, or in fact, any decidable logic beyond NP.

The Exponential Time Hypothesis (ETH) was first introduced and studied by Implagiazzo, Paturi and Zane [17, 18]. Briefly, it states that for every $k \geqslant 3$, there is a constant $c_k > 1$ such that any (deterministic/randomized) algorithm that decides $k$-SAT has run time $\Omega(c_k^n)$, where $n$ is the number of variables. Essentially it states that there is no subexponential algorithm that decides $k$-SAT, which is a stronger statement than P $\neq$ NP. The stronger version of ETH, called the *Strong Exponential Time Hypothesis* (SETH) introduced by Calabro, Implagiazzo and Paturi [5] states that the sequence $c_3, c_4, \ldots$ converges to 2, i.e., $\lim_{k\to\infty} c_k = 2$. Intuitively, it states that any algorithm that decides SAT has run time $\Omega(2^n)$. Since its introduction, ETH as well as SETH have often been used as the standard tools to establish the lower bounds of various NP-complete problems. See, e.g., [7, 31] and the references within for more details.

Galperin and Wigderson [10] proposed and studied a notion of succinctly represented graphs, in which graphs are represented as Boolean circuits, instead of, as lists of their edges. Papadimitriou and Yannakakis [30] showed that in such representation many NP-complete graph-theoretic problems become NEXP-complete. This notion of succinct representation is quite clearly different from the way SAT($\mathsf{FO}^2$) is a succinct representation of CIS.

Finally, we note that there have been work where various extensions of two-variable logic are reduced to graph theoretic problems. See, e.g., [3, 21, 23]. However, the techniques are different from ours and they do not imply any efficient algorithms for SAT($\mathsf{FO}^2$).

*Organization.* This paper is organized as follows. We present the formal definition of CIS and its algorithms in Section 2 where we also identify a few tractable fragments of CIS. In Section 3 we consider SAT($\mathsf{FO}^2$) when the equality predicate is absent, and identify fragments of $\mathsf{FO}^2$ that parallel the tractable fragments of CIS. In Section 4 we present the reduction from SAT($\mathsf{FO}^2$) with the equality predicate to the fragment without the equality predicate. We present some of our initial experimental results in Section 5. Finally, we conclude in Section 6.

## ACKNOWLEDGMENTS

The preliminary version of this paper has appeared in the Proceedings of the ACM/IEEE 36th annual symposium on Logic in Computer Science (LICS) 2021. In that paper we propose three algorithms where the third one is obtained by a slight modification of the second one and we claimed its run time is $O(1.3661^n)$. It turns out that its analysis contains wrong calculation. The algorithm is correct, but the correct calculation yields the same run time as the second one. So we decided to omit the third algorithm in this journal version. Instead we include a new lower bound for CIS. This paper has also been published in arXiv [26] which contains more experimental results along with the detailed descriptions of the benchmarks.

We would like to thank Phokion Kolaitis for his suggestion to consider the connection between ETH/SETH and CIS. We also would like to thank Bartosz Bednarczyck, Michael Benedikt, Filip Murlak and Anthony Widjaja Lin for their constructive suggestions and feedback, as well as the anonymous LICS reviewers for their thorough review. We acknowledge the generous financial support from Taiwan Ministry of Science and Technology grant no. 109-2221-E-002-143-MY3 and National Taiwan University grant no. 109L891808.

## 2 CONDITIONAL INDEPENDENT SET

We divide this section into four subsections. Subsection 2.1 contains the formal definition of CIS and terminology that we will use in this paper. We show that CIS is NP-complete. Then, in Subsection 2.2 and 2.3 we describe our deterministic and randomized algorithms for CIS. Finally, we present some tractable fragments of CIS in Subsection 2.5.

## 2.1 Definition and terminology

A *graph system* is a tuple $\mathcal{G} = (G_0, G_1, \ldots, G_m)$, with $m \geqslant 1$, where $G_0, G_1, \ldots, G_m$ are all graphs over the same (finite) set of vertices, denoted by $V(\mathcal{G})$, but $G_0$ is an undirected graph and all the others $G_1, \ldots, G_m$ are directed graphs. We denote by $E_i(\mathcal{G})$ the set of edges in graph $G_i$, for each $0 \leqslant i \leqslant m$.

We follow the convention of writing an edge as a pair $(u, v)$ of vertices. However, an edge $(u, v) \in E_i(\mathcal{G})$, where $1 \leqslant i \leqslant m$, is understood to be a directed edge that goes from $u$ to $v$. For technical reason that will become apparent later, we assume $G_0$ does not contain self-loops, whereas $G_1, \ldots, G_m$ may contain self-loops. To avoid clutter, we write $V$ and $E_i$, instead of $V(\mathcal{G})$ and $E_i(\mathcal{G})$, when $\mathcal{G}$ is already clear from the context.

We call $G_0$ the *conflict graph* in $\mathcal{G}$. Two adjacent vertices in $G_0$ are called *conflicting* vertices. An independent set in $\mathcal{G}$ is an independent set in $G_0$, i.e., a set that does not contain two conflicting vertices.

For a set $\Gamma \subseteq V$, for $1 \leqslant i \leqslant m$, we say that a vertex $u \in \Gamma$ is a $G_i$*-good* vertex in $\Gamma$, if there is $v \in \Gamma$ such that $(u, v) \in E_i$. Note that $E_i$ may contain self-loops, thus, if $(u, u) \in E_i$, $u$ is $G_i$-good in $\Gamma$. If $u$ is $G_i$-good in $\Gamma$ for every $1 \leqslant i \leqslant m$, we call it a *good* vertex in $\Gamma$. Otherwise, it is called a *bad* vertex in $\Gamma$. A set $\Gamma$ is a *good independent set* (GIS) in $\mathcal{G}$, if it is a non-empty independent set and each of its vertices is a good vertex in $\Gamma$.

Intuitively, one may view the graph $G_i$, where $1 \leqslant i \leqslant m$, as a kind of dependency graph depicting a condition that "a vertex may be picked only if one of its outgoing neighbours in $G_i$ is picked." Thus, a set $\Gamma$ is a GIS, if it does not contain any two conflicting vertices and for every vertex in $\Gamma$, at least one of its outgoing neighbours in $G_i$ is also in $\Gamma$, for every $1 \leqslant i \leqslant m$.

We define the problem *Conditional Independent Set* (CIS) as given a graph system $\mathcal{G}$, decide if it has a GIS. In language theoretic term, CIS is the set $\{\mathcal{G} \mid \mathcal{G} \text{ has a GIS}\}$. We show that it is NP-complete, as stated below.

Theorem 2.1. *CIS is NP-complete and it is already NP-hard when $m = 1$.*

Proof. The NP membership is straightforward. The hardness is obtained by reduction from the independent set problem: Given an undirected graph $G = (V, E)$ and an integer $k \leqslant |V|$, decide if $G$ has an independent set of size $k$.

We construct a graph system $\mathcal{G} = (G_0, G_1)$ where the set of vertices is $V(\mathcal{G}) = V \times \{0, \ldots, k-1\}$. The set of edges in $G_0$ is as follows.

- For every $0 \leqslant i \leqslant k-1$, the set $V \times \{i\}$ forms a clique in $G_0$.
- For every vertex $u \in V$, the set $\{u\} \times \{0, \ldots, k-1\}$ forms a clique in $G_0$.
- For every $0 \leqslant i \neq j \leqslant k-1$, for every edge $(u, v) \in E$, $((u, i), (v, j))$ is an edge in $G_0$.

The set of edges in $G_1$ is as follows. For every $0 \leqslant i \leqslant k-1$, there are directed edges from every vertex in $V \times \{i\}$ to every vertex in $V \times \{i + 1 \pmod{k}\}$.

It is routine to show that $G$ has an independent set of size $k$ if and only if $\mathcal{G}$ has a GIS.

(only if) Let $I = \{v_0, \ldots, v_{k-1}\}$ be an independent set of size $k$. It is not difficult to see that the set $\{(v_0, 0), \ldots, (v_{k-1}, k-1)\}$ is a GIS in $\mathcal{G}$.

(if) Let $\Gamma$ be a GIS in $\mathcal{G}$. For $0 \leqslant i \leqslant k-1$, let $V_i = V \times \{i\}$. By definition of the edges in $G_1$, if there is $0 \leqslant i \leqslant k-1$ such that $\Gamma \cap V_i \neq \emptyset$, then $\Gamma \cap V_{i+1 \pmod{k}} \neq \emptyset$. Thus, for every $0 \leqslant i \leqslant k-1$, $\Gamma \cap V_i \neq \emptyset$. Moreover, since each $V_i$ forms a clique in $G_0$, $\Gamma$ contain exactly one vertex from each $V_i$. Let $(v_0, 0), \ldots, (v_{k-1}, k-1)$ be the vertices in $\Gamma$. By the definition of edges in $G_0$, the set $\{v_0, \ldots, v_{k-1}\}$ is an independent set in $\mathcal{G}$. □

## 2.2 The first algorithm for CIS

It is straightforward to design a deterministic algorithm for CIS with run time $O(2^n \cdot \mathrm{poly}(n))$. In this subsection we will present our deterministic algorithm for CIS with significantly lower complexity, i.e., $O(\delta_1^n)$, where $\delta_1 = 1.4423 \approx \sqrt[3]{3}$.

We start with a simple Procedure 1 below that on input a graph system $\mathcal{G}$ and an independent set $Y$ in $\mathcal{G}$, decides if $\mathcal{G}$ has a GIS $\Gamma$ where $\Gamma \subseteq Y$.

**Procedure 1**

**Input:** A graph system $\mathcal{G}$ and an independent set $Y$ in $\mathcal{G}$.
**Task:** Return true if and only if $\mathcal{G}$ has a GIS $\Gamma$ where $\Gamma \subseteq Y$.

```
1: Γ := Y.
2: while there is a bad vertex u in Γ do
3:     Γ := Γ \ {u}.
4: if Γ ≠ ∅ then
5:     return true.
6: else
7:     return false.
```

Procedure 1 runs in polynomial time, since checking whether a set contains a bad vertex can be done in polynomial time. To prove correctness, let $Y$ be an independent set. Note that if there are two GIS $\Gamma_1$ and $\Gamma_2$ in $Y$, their union $\Gamma_1 \cup \Gamma_2$ is also a GIS in $Y$. This means that if there is a GIS in $Y$, then there is a unique maximal GIS in $Y$. Hence, the **while**-loop will iterate until $\Gamma$ becomes the maximal GIS, and Procedure 1 returns true. On the other hand, if there is no GIS in $Y$, every subset of $Y$ contains a bad vertex. Hence, the **while**-loop will iterate until $\Gamma$ becomes $\emptyset$ and Procedure 1 returns false.

We use Procedure 1 to design our first algorithm for CIS, presented as Algorithm-A below.

Algorithm-A

**Input:** A graph system $\mathcal{G}$.
**Task:** Return true if and only if $\mathcal{G}$ has a GIS.

```
1: for all maximal independent set Y do
2:     Using Procedure 1, decide if there is a GIS Γ ⊆ Y.
3:     if there is such a GIS then
4:         return true.
5: return false.
```

Its correctness is immediate from Procedure 1 since any GIS is contained inside some maximal independent set. A well known result of Moon and Moser [28] states that there are at most $O(3^{n/3})$ maximal independent sets in a graph of $n$ vertices. Moreover, there are algorithms [4, 42, 43] that lists all those sets in $O(3^{n/3} \cdot \mathrm{poly}(n))$ time. By rounding up $\sqrt[3]{3}$ to $\delta_0 = 1.4423$, Algorithm-A runs in $O(\delta_0^n)$ time. We state this formally as Theorem 2.2.

Theorem 2.2. *Algorithm-A decides CIS in time $O(\delta_0^n)$ time, where $n$ is the number of vertices and $\delta_0 = 1.4423 \approx \sqrt[3]{3}$.*

## 2.3 The second algorithm for CIS

Next, we present our randomized algorithms. We start with the following remark.

Remark 1. Let $\mathcal{G}$ be a graph system and let $X \subseteq Y \subseteq V(\mathcal{G})$. Let $u$ and $v$ be two vertices in $Y$ that are conflicting. Suppose there is GIS $\Gamma$ such that $X \subseteq \Gamma \subseteq Y$. Then, at least one of the following holds:

$$X \subseteq \Gamma \subseteq Y \setminus \{u\} \qquad \text{or} \qquad X \cup \{u\} \subseteq \Gamma \subseteq Y \setminus \{v\}.$$

Indeed, if $u \in \Gamma$, then $v \notin \Gamma$, implying the second case. If $u \notin \Gamma$, the first case holds trivially.

Remark 1 immediately gives us a recursive procedure, presented as Procedure Las-Vegas below, for checking if there is a GIS in between two sets $X$ and $Y$.

**Procedure** Las-Vegas

**Input:** A graph system $\mathcal{G}$ and two sets $X, Y \subseteq V(\mathcal{G})$ where $X \subseteq Y$ and $X$ and $Y$ are not necessarily independent.

**Task:** Return true if and only if $\mathcal{G}$ has a GIS in between $X$ and $Y$.

```
if X is not independent set, return false.
Remove all vertices in Y that are conflicting with some vertex in X.
Remove all bad vertices from Y (like in Step 2 in Procedure 1).
if Y ⊊ X or Y = ∅, return false.
if Y is independent set, return true.
{Note that if this line is reached, there are at least two conflicting vertices in Y \ X.}
Let u, v ∈ Y be two conflicting vertices in Y.
if Las-Vegas(G, X, Y \ {u}) then
   return true.
else
   return Las-Vegas(G, X ∪ {u}, Y \ {v}).
```

Note that Las-Vegas can be easily turned into a randomized algorithm with zero error, where for every conflicting vertices $(u, v)$, we randomly choose one to be omitted first. Indeed this is what we implement in the experiment section. Our second algorithm for CIS, presented as Algorithm-B, runs Las-Vegas with $X = \emptyset$ and $Y = V(\mathcal{G})$. Its correctness and analysis is stated in Theorem 2.3.

Algorithm-B:

**Input:** A graph system $\mathcal{G}$.

**Task:** Return true if and only if $\mathcal{G}$ has a GIS.

```
return Las-Vegas(G, ∅, V(G)).
```

Theorem 2.3. Algorithm-B *decides CIS with run time $O(\delta_1^n)$, where n is the number of vertices and $\delta_1 = 1.6181 \approx (\sqrt{5}+1)/2$.*

Proof. Correctness is immediate from Las-Vegas and Remark 1. For the run time analysis, let $T_{j,n}$ denote the run time of Las-Vegas, where $j = |Y| - |X|$ and $n$ is the number of vertices. Each $T_{j,n}$ can be defined by the following recurrence relation:

$$\begin{aligned} T_{0,n} &= T_{1,n} = O(n^2) \\ T_{j,n} &\leqslant T_{j-1,n} + T_{j-2,n} + O(n^2) \qquad (\text{for } j \geqslant 2) \end{aligned}$$

Note that $T_{j,n} = O(F_j \cdot n^2)$, where $F_j$ is the $j$-th Fibonacci number. By the generating function method [13, 44], $T_{j,n} = O(\delta_1^j)$, for every $j \geqslant 0$. Thus, the run time of Algorithm-B is $T_{n,n} = O(\delta_1^n)$. □

## 2.4 A lower bound for CIS

In this section we will present a lower bound for CIS assuming that the so called *(Strong) Exponential Time Hypothesis* holds [5, 17, 18]. We first present a brief review. For an integer $k \geqslant 3$, let $c_k$ be the infimum of the real numbers $\varepsilon$ such that any (deterministic/randomized) algorithm that decides $k$-SAT runs in time $O(\varepsilon^n)$, where $n$ is the number of variables. The *Exponential Time Hypothesis* (ETH) states that $c_k > 1$, for every $k \geqslant 3$. The sequence $c_3, c_4, c_5, \ldots$ is monotone and bounded by 2. The *Strong Exponential Time Hypothesis* (SETH) states that the limit $\lim_{k\to\infty} c_k = 2$. Intuitively, SETH states that every algorithm that decides SAT runs in time $\Omega(2^n)$.

Using SETH, we obtain a lower bound for CIS as stated in Theorem 2.4.

Theorem 2.4. *Unless SETH fails, every algorithm that decides CIS has run time $\Omega(\delta_2^n)$, where $n$ is the number of vertices and $\delta_2 = \sqrt{2}$.*

Proof. We will show that every $k$-CNF formula $\varphi$ with $n$ variables can be reduced to an instance of CIS with $2n+1$ vertices. Let $\varphi$ be a $k$-CNF formula with variables $x_1, \ldots, x_n$ and clauses $C_1, \ldots, C_m$. We construct the following graph system $\mathcal{G} = (G_0, G_1, \ldots, G_{n+m+1})$.

(P1) It has $2n + 1$ vertices: $z, x_1, \ldots, x_n, \neg x_1, \ldots, \neg x_n$.
(P2) The graph $G_0$ has $n$ edges: $(x_1, \neg x_1), \ldots, (x_n, \neg x_n)$, i.e., each $x_i$ and $\neg x_i$ are conflicting.
(P3) For every $i = 1, \ldots, n$, the graph $G_i$ has the following edges:
- $(z, x_i)$ and $(z, \neg x_i)$.
- For every literal $\ell$, it has self-loop $(\ell, \ell)$.

(P4) For every $i = 1, \ldots, m$, the graph $G_{n+i}$ has the following edges.
- It has edge $(z, \ell)$, if the clause $C_i$ contains literal $\ell$.
- For every literal $\ell$, it has self-loop $(\ell, \ell)$.

(P5) The graph $G_{n+m+1}$ contains the edge $(\ell, z)$, for every literal $\ell$.

We observe that that $\varphi$ is satisfiable if and only if $\mathcal{G}$ has a GIS.

(only if) Let $\xi$ be a satisfying assignment for $\varphi$. We think of $\xi$ such that either $x_i$ or $\neg x_i$ is in $\xi$, for every $i = 1, \ldots, n$. Then, the set $\Gamma = \xi \cup \{z\}$ is a GIS in $\mathcal{G}$.

(if) Suppose $\Gamma$ is a GIS in $\mathcal{G}$. Due to (P5), vertex $z$ must be in $\Gamma$. Due to (P2) and (P3), $\Gamma \cap \{x_1, \ldots, x_n, \neg x_1, \ldots, \neg x_n\}$ is a set $\xi$ such that either $x_i$ or $\neg x_i$ is in $\xi$, for every $i = 1, \ldots, n$. Thus, it is a boolean assignment for the variables $x_1, \ldots, x_n$. In fact, by (P4), every clause is satisfied by the assignment $\xi$. Thus, $\varphi$ is satisfiable.

To complete the proof of Theorem 2.4, suppose we have an algorithm that decides CIS in time $O(c^n)$. The reduction above implies SAT can be decided in time $O(c^{2n+1}) = O(c^{2n})$, which by SETH, $c^2 \geqslant 2$, i.e., $c \geqslant \sqrt{2}$. □

## 2.5 Tractable fragments

In this subsection we will present a few tractable fragments of CIS with complexity Ptime-complete, NLog-complete and DLog, respectively. We start with the following terminology.

*Definition 2.5.* Let $\mathcal{G} = (G_0, G_1, \ldots, G_m)$.

- We say that $\mathcal{G}$ is *conflict-free*, if $G_0$ does not contain any edge, i.e., it does not contain any conflicting vertices.
- We say that $\mathcal{G}$ is *uniquely-outgoing*, if for every $1 \leqslant i \leqslant m$, every vertex has at most one outgoing edge in $G_i$.

*Ptime-complete fragment.* We first show that CIS drops to Ptime-complete on conflict-free graph systems. Note that Ptime membership is immediate, since both Algorithm-A and -B run in polynomial time on conflict-free graph systems. In the case of Algorithm-A, there is only one maximal independent set, while in the case of Algorithm-B, there is no recursive call.

Hardness is obtained by log-space reduction from the reachability problem for alternating graphs, which is known to be Ptime-complete [16, Theorem 3.26]. We first present a brief review of some definitions from [16]. An *alternating* graph is a directed graph $G = (V, E, A)$, where $A \subseteq V$. Vertices in $A$ and $V \setminus A$ are called *universal* and *existential* vertices, respectively. We assume that all universal vertices have at least one outgoing edge. For two vertices $s, t \in V$, we say that *$t$ is reachable from $s$ in $G$*, if there is a rooted tree $T$ whose nodes are labeled with vertices from $G$ such that the following holds.

- The root node is labeled with $s$.
- All the leaf nodes in $T$ are labeled with $t$.
- For every non-leaf node $x$ in $T$, the following holds.
  - If $x$ is labeled with an existential vertex $u \in V \setminus A$, then $x$ has only one child labeled with vertex $v$ where $(u, v) \in E$.
  - If $x$ is labeled with a universal vertex $u \in A$ with $k$ outgoing edges $(u, v_1), \ldots, (u, v_k)$, then $x$ has $k$ children $z_1, \ldots, z_k$ labeled with $v_1, \ldots, v_k$, respectively.

The reachability problem for alternating graphs is defined as follows. On input alternating graph $G$ and two vertices $s, t$, decide if $t$ is reachable from $s$ in $G$. It is not difficult to see that this problem is just a graph theoretic formulation of alternating logarithmic space Turing machines, which are equivalent with deterministic polynomial time Turing machines [6].

Now we present a log-space reduction from the reachability problem for alternating graphs to the reachability problem in alternating graphs. Let $G = (V, E, A)$ be an alternating graph, where $A \subseteq V$ is the set of universal vertices and let $s, t \in V$. Without loss of generality, we may assume that every universal vertex has exactly 2 outgoing edges and that $t$ is an existential vertex without outgoing edge.

We construct the following graph system $\mathcal{G} = (G_0, G_1, G_2)$. The set of vertices is $V(\mathcal{G}) = V \times \{1, \ldots, n\}$, where $n = |V|$. Graph $G_0$ does not contain any edge. The set of edges in $G_1$ and $G_2$ are as follows.

- $((t, n), (s, 1))$ and $((t, i), (t, i + 1))$ are edges in both $G_1$ and $G_2$, for every $1 \leqslant i \leqslant n - 1$.
- For every edge $(u, v)$ in $G$, where $u$ is an existential vertex, i.e., $u \notin A$, $((u, i), (v, i + 1))$ is an edge in $G_1$, for every $1 \leqslant i \leqslant n - 1$.
- For every vertex $u \in V \setminus A$, the self loop $((u, i), (u, i))$ is an edge in $G_2$, for every $1 \leqslant i \leqslant n$.
- For every universal vertex $u \in A$ with outgoing edges $(u, v_1)$ and $(u, v_2)$ in $G$, $((u, i), (v_1, i+1))$ is an edge in $G_1$ and $((u, i), (v_2, i + 1))$ is an edge in $G_2$, for every $1 \leqslant i \leqslant n - 1$.

It is straightforward that the reduction can be done in log space. The correctness follows from Lemma 2.6 below.

Lemma 2.6. *$\mathcal{G}$ has a GIS if and only if vertex $t$ is reachable in $G$ from vertex $s$.*

Proof. (if) Suppose $t$ is reachable from $s$ in $G$. Let $T$ be the witness tree and let $d$ be the depth. We define the depth of the root node of $T$ as 1. Obviously we can also assume that $d \leqslant n$.

Let $\Gamma \subseteq V(\mathcal{G})$ be the following set.

$$\Gamma \quad := \quad \{(t, 1), \ldots, (t, n)\} \ \cup \ \{(u, i) \mid \text{there is a non-leaf node in } T \text{ with label } u \text{ and depth } i\}$$

We claim that every vertex $(u, i) \in \Gamma$ is a good vertex in $\Gamma$, and hence, $\Gamma$ is a GIS in $\mathcal{G}$.

First, $(t, n)$ is not a bad vertex, since by definition, $(s, 1) \in \Gamma$ and $((t, n), (s, 1))$ is an edge in both $G_1$ and $G_2$. Similarly, all vertices $(t, 1), \ldots, (t, n - 1)$ are not bad vertices.

Let $(u, i) \in \Gamma$, where $u \neq t$. Thus, there is a non-leaf node $x$ in $T$ with label $u$. There are two cases.

- $u$ is an existential vertex in $G$.
By construction of $\mathcal{G}$, there is a self-loop $((u, i), (u, i))$ in $G_2$. Therefore, $(u, i)$ is $G_2$-good. We now show that $(u, i)$ is $G_1$-good. By definition, $x$ has only one child. Let $v$ be the label of this child, thus, $(u, v) \in E$. If $v$ is a leaf node, then $v = t$. Since $(t, i+1) \in \Gamma$ and $((u, i), (t, i+1))$ is an edge in $G_1$, the vertex $(u, i)$ is $G_1$-good. Similarly, if $v$ is not a leaf node, $v$ has depth $i + 1$. By definition of $\Gamma$, $((u, i), (v, i + 1))$ is an edge in $G_1$. Thus, $(u, i)$ is $G_1$-good.
- $u$ is a universal vertex in $G$.
By definition, $u$ has two outgoing edges in $G$, denoted by $(u, v_1)$ and $(u, v_2)$. By the construction of $\mathcal{G}$, we can assume that $G_1$ contains the edges $((u, i), (v_1, i + 1))$ and $G_2$ contains the edges $((u, i), (v_2, i + 1))$.
Now, $x$ has two children $y_1$ and $y_2$, labeled with $v_1$ and $v_2$, respectively. If $y_1$ is a leaf node, then $v_1 = t$. Since $(t, i + 1) \in \Gamma$ and $((u, i), (t, i + 1))$ is an edge in $G_1$, $(u, i)$ is $G_1$-good. If $y_1$ is not a leaf node, then $(v_1, i + 1) \in \Gamma$. Since $((u, i), (v_1, i + 1))$ is an edge in $G_1$, $(u, i)$ is $G_1$-good. The proof that $(u, i)$ is $G_2$-good is similar, thus, omitted.

(only if) Suppose $\mathcal{G}$ has a GIS $\Gamma$. Note that edges in $G_1$ are only one directional, i.e., they can only go from $(u, i)$ to $(v, i + 1)$. The only exception is the edge $((t, n), (s, 1))$. So, for every vertex in $\Gamma$ to be $G_1$-good, the set $\Gamma$ must contain both $(t, n)$ and $(s, 1)$. We have the following claim that immediately implies that $t$ is reachable from $s$ in $G$.

Claim 1. *For every vertex $(u, i) \in \Gamma$, $t$ is reachable from $u$ in $G$.*

Proof. (of claim) The proof is by "backward" induction on $i$.
The base case is $i = n$. Since $(u, n)$ is $G_1$-good in $\Gamma$, vertex $(u, n)$ must be $(t, n)$. The claim holds trivially, since $t$ is reachable from $t$ in $G$.

For the induction hypothesis, we assume that for every $(u, i) \in \Gamma$, $t$ is reachable from $u$ in $G$. The induction step is as follows. Let $(u, i - 1) \in \Gamma$. There are two cases.

- $u$ is an existential vertex in $G$.
Since $(u, i - 1)$ is $G_1$-good, there is some $(v, i) \in \Gamma$ such that $((u, i - 1), (v, i))$ is an edge in $G_1$. Applying induction hypothesis on $(v, i)$, $t$ is reachable from $v$ in $G$. Let $T$ be the witness tree. Since $((u, i - 1), (v, i))$ is an edge in $G_1$, $(u, v)$ is an edge in $G$. Thus, $t$ is reachable from $u$ in $G$, where the witness tree $T'$ is obtained by by inserting a new root node $x$ with label $u$, and the child of $x$ is the root node of $T$.
- $u$ is a universal vertex in $G$.
Since $(u, i - 1)$ is $G_1$- and $G_2$-good, there is some $(v_1, i), (v_2, i) \in \Gamma$ and $((u, i - 1), (v_1, i))$ and $((u, i - 1), (v_2, i))$ are edges in $G_1$ and $G_2$, respectively. Applying induction hypothesis on $(v_1, i)$ and $(v_2, i)$, $t$ is reachable from both $v_1$ and $v_2$ in $G$. Let $T_1$ and $T_2$ be the respective witness trees.
Since $((u, i - 1), (v_1, i))$ and $((u, i - 1), (v_2, i))$ are edges in $G_1$ and $G_2$, respectively, $(u, v_1)$ and $(u, v_2)$ are edges in $G$. Now, $t$ is reachable from $u$ in $G$, where the witness tree $T'$ is obtained by by inserting a new root node $x$ with label $u$, and $x$ has two children which are the root nodes of $T_1$ and $T_2$.

□

This completes our proof of Lemma 2.6. □

*NLog-complete fragment.* Next, we show that CIS drops to NLog-complete when restricted to conflict-free graph systems $\mathcal{G} = (G_0, G_1)$, i.e., $m = 1$. Indeed in this case CIS is equivalent to checking the existence of a cycle in a directed graph, as stated in Lemma 2.7 below. Note that it is a folklore that checking the existence of a cycle in a directed graph is NLog-complete.

Lemma 2.7. *For every conflict-free graph system $\mathcal{G} = (G_0, G_1)$, $\mathcal{G}$ has a GIS if and only if $G_1$ contains a cycle.*

Proof. (if) A cycle in $G_1$ does not contain any bad vertex, and hence, it is a GIS. (only if) Let $\Gamma$ be a GIS. By definition, every vertex in $\Gamma$ has an outgoing neighbor in $\Gamma$. Since $\Gamma$ is finite, this implies that there is a cycle in $\Gamma$. □

*Uniquely-outgoing graph systems.* Finally, we show that restricted to uniquely outgoing graph systems with $m \geqslant 2$, CIS is NLog-complete. When $m = 1$, it drops to deterministic log-space.

Let $\mathcal{G} = (G_0, G_1, \dots, G_m)$ be uniquely-outgoing. For a vertex $u$, let $\mathcal{R}(u)$ denote the set of vertices $v$ such that there is a path $w_1, \dots, w_k$, where $u = w_1$, $v = w_k$ and each $(w_j, w_{j+1}) \in \bigcup_{i=1}^{m} E_i(\mathcal{G})$. Intuitively, $\mathcal{R}(u)$ is the set of vertices reachable from vertex $u$ using edges in $\bigcup_{i=1}^{m} E_i(\mathcal{G})$.

We have the following characterization of the existence of GIS for uniquely-outgoing graph systems.

Lemma 2.8. *If a uniquely-outgoing graph system $\mathcal{G}$ has a GIS, then there is a vertex $u \in V(\mathcal{G})$ such that $\mathcal{R}(u)$ is GIS in $\mathcal{G}$.*

Proof. Let $\Gamma$ be a GIS in $\mathcal{G}$ and $u \in \Gamma$. Since $\mathcal{G}$ is uniquely outgoing, $\mathcal{R}(u) \subseteq \Gamma$. Otherwise, $\Gamma$ contains a bad vertex. This also implies that $\mathcal{R}(u)$ does not contain conflicting vertices, since $\mathcal{R}(u) \subseteq \Gamma$. Therefore, $\mathcal{R}(u)$ is also a GIS. □

We will now show that CIS is NLog-complete, when restricted to uniquely-outgoing graph systems. For the upper bound, since NLog is closed under complement [15, 41], we describe a non-deterministic log-space algorithm that decides whether a uniquely outgoing graph system does *not* have a GIS. It works as follows. Let $\mathcal{G} = (G_0, G_1, \dots, G_m)$ be the input. Iterate through every vertex $u \in V(\mathcal{G})$ and check if one of the following hold.

(a) There is $1 \leqslant i \leqslant m$ and a vertex $v \in \mathcal{R}(u)$ that does not have outgoing edge in $G_i$.
(b) $\mathcal{R}(u)$ contains conflicting vertices.

To check (a), guess a path from $u$ to some vertex $v$ that does not have an outgoing edge in $G_i$ for some $1 \leqslant i \leqslant m$. To check (b), guess two paths from $u$ to two vertices $v_1$ and $v_2$ and verify that $v_1, v_2$ are conflicting vertices. It is straightforward that this algorithm uses only logarithmic space.

For the lower bound, note that non-reachability problem for standard directed graphs can be expressed as non-reachability problem for alternating graphs where all vertices are universal vertices. In this case, our reduction (for Ptime-hardness above) will yield uniquely-outgoing graph systems. (See the fourth bullet in the construction of $\mathcal{G} = (G_0, G_1, G_2)$ in our reduction for Ptime-hardness above.) Since NLog is closed under complement, NLog-hardness follows immediately.

Here we remark that for uniquely-outgoing $\mathcal{G} = (G_0, G_1)$, i.e., $m = 1$, every vertex $v \in \mathcal{R}(u)$ is reachable from $u$ by a unique path. Thus, our algorithm for checking (a) above works deterministically, since every vertex has at most one outgoing edge. To check (b), we can do the following: For every pair $(v_1, v_2)$ of conflicting vertices, we check whether $v_1$ and $v_2$ are both reachable from $u$. Therefore, restricted to uniquely-outgoing graph systems with $m = 1$, CIS is decidable in logarithmic space.

We summarize formally the results in this subsection as Theorem 2.9 below.

Theorem 2.9.

- *CIS is Ptime-complete on conflict-free graph systems with $m \geqslant 2$.*
- *CIS is NLog-complete on conflict-free graph systems with $m = 1$.*
- *CIS is NLog-complete on uniquely-outgoing graph systems with $m \geqslant 2$.*
- *CIS is decidable in deterministic log-space on uniquely-outgoing graph systems with $m = 1$.*

## 3 $FO^2$ WITHOUT THE EQUALITY PREDICATE

As mentioned earlier, $FO^2$ denotes the fragment of relational first-order logic that uses only two variables: $x$ and $y$. For convenience, we assume that only unary and binary predicates are used. In this section we focus on $FO^2$ without the equality predicate. We introduce some standard terminology in Subsection 3.1. In Subsection 3.2 we show how to reduce SAT($FO^2$) to CIS. Finally, in Subsection 3.3 we present the fragments of $FO^2$ that correspond exactly to the tractable fragments of CIS.

### 3.1 Terminology

We first recall a well known result by Scott that every $FO^2$ sentence can be rewritten in linear time (over an extended vocabulary) into Scott normal form [40]:

$$\Phi \quad := \quad \forall x \forall y \; \alpha(x,y) \;\wedge\; \bigwedge_{i=1}^{m} \forall x \exists y \; \beta_i(x,y) \tag{1}$$

where each $\beta_i(x,y)$ and $\alpha(x,y)$ are all quantifier free formulas. If the original sentence does not contain the equality predicate, neither do the formulas $\beta_i(x,y)$ and $\alpha(x,y)$. We also note that the transformation to Scott normal form introduces new predicates, thus, it only yields equisatisfiable formulas.

For the rest of this section, let $n$ be the total number of predicates used in $\Phi$. We recall a few standard terminologies. A *unary literal* (in short, 1-literal) is an atomic predicate or its negation using only variable $x$, and a *binary literal* (in short, 2-literal) is an atomic predicate or its negation using both variables $x$ and $y$. A *literal* is either a 1- or 2-literal. Note that atom like $R(x,x)$, where $R$ is a binary relation, is 1-literal. A 2-literal is always of the form $R(x,y)$ or $R(y,x)$ or their negations.

A *unary type* (in short, 1-type) is defined as a maximally consistent set of unary literals and a *binary type* (in short, 2-type) is a maximally consistent set of binary literals. A *type* is either a 1- or 2-type. Note that a type can be viewed as a quantifier-free formula that is the conjunction of its elements. Alternatively, it can also be viewed as partial Boolean assignment to atomic predicates. The number of 1- and 2-types are at most $2^n$ and $2^{4k}$, respectively, where $k$ is the number of binary predicates. We will use the symbols $\pi$ and $\eta$ (possibly indexed) to denote 1-type and 2-type, respectively. When viewed as formula, we write $\pi(x)$ and $\eta(x,y)$, respectively. We write $\pi(y)$ to denote the formula $\pi(x)$ with $x$ being substituted with $y$.

For a structure $\mathcal{A}$, the *type of an element* $a \in A$ is the unique 1-type $\pi$ that $a$ satisfies in $\mathcal{A}$. Similarly, the type of a pair $(a,b) \in A \times A$ is the unique 2-type that $(a,b)$ satisfies in $\mathcal{A}$. We say that a 1- or 2-type is *realized* in $\mathcal{A}$ if there is an element/a pair of elements that satisfies it.

### 3.2 Reduction to CIS

Let $\Phi$ be as in Eq. (1). For a 1-type $\pi$, we say that $\pi$ *is compatible with* $\alpha(x,y)$, if there is a 2-type $\eta$ such that $\pi(x) \wedge \eta(x,y) \wedge \pi(y) \models \alpha(x,y)$. Likewise, for two 1-types $\pi_1$ and $\pi_2$ (not necessarily different), we say that $(\pi_1,\pi_2)$ *is compatible with* $\alpha$, if there is a 2-type $\eta$ such that $\pi_1(x) \wedge \eta(x,y) \wedge \pi_2(y) \models \alpha(x,y)$. Otherwise, we say that $(\pi_1,\pi_2)$ *is incompatible with* $\alpha$. For some $1 \leqslant i \leqslant m$, we say that $(\pi_1,\pi_2)$ *is* $\beta_i$*-compatible with* $\alpha$, if there is 2-type $\eta$ such that $\pi_1(x) \wedge \eta(x,y) \wedge \pi_2(y) \models \alpha(x,y) \wedge \beta_i(x,y)$.

Now, let $\mathcal{G}_\Phi = (G_0, G_1, \ldots, G_m)$ be a graph system defined as follows.

- The set $V(\mathcal{G})$ of vertices is the set of 1-types that are compatible with $\alpha$.
- The graph $G_0$ consists of the following edges: $(\pi_1,\pi_2)$ is an edge if and only if $(\pi_1,\pi_2)$ is incompatible with $\alpha$.

- For each $1 \leqslant i \leqslant m$, the graph $G_i$ consists of the following edges: $(\pi_1, \pi_2)$ is an edge if and only if $(\pi_1, \pi_2)$ is $\beta_i$-compatible with $\alpha$.

Note that $\pi_1(x) \wedge \eta(x, y) \wedge \pi_2(y)$ is a full Boolean assignment to the atomic predicates in $\alpha(x, y)$. Thus, deciding whether $\pi$ or $(\pi_1, \pi_2)$ is compatible or $\beta_i$-compatible with $\alpha$ is essentially Boolean SAT problem. Therefore, the construction of the graph system $\mathcal{G}_\Phi$ takes exponential time in $n$.

Next, we prove Lemma 3.1 that links SAT(FO$^2$) with CIS.

Lemma 3.1. *$\Phi$ is satisfiable if and only if $\mathcal{G}_\Phi$ has a GIS.*

Proof. (only if) Let $\mathcal{A} \models \Phi$. Obviously, every 1-type realized in $\mathcal{A}$ is compatible with $\alpha$. Consider the set $\Gamma := \{\pi | \pi \text{ is realized in } \mathcal{A}\}$, which is a subset of $V(\mathcal{G}_\Phi)$. We claim that $\Gamma$ is a GIS in $\mathcal{G}_\Phi$. First, we show that $\Gamma$ is an independent set. Let $\pi_1, \pi_2 \in \Gamma$. By definition, there is $a, b \in A$ whose 1-types are $\pi_1$ and $\pi_2$, respectively. Since $\mathcal{A} \models \forall x \forall y\ \alpha(x, y)$, we have:

$$\mathcal{A}, x/a, y/b \models \alpha(x, y) \quad \text{and} \quad \mathcal{A}, x/b, y/a \models \alpha(x, y)$$

This means that $(\pi_1, \pi_2)$ are not incompatible with $\alpha$. Thus, $\pi_1$ and $\pi_2$ are not adjacent in the graph $G_0$.

Next, we show that for every $1 \leqslant i \leqslant m$, $\Gamma$ is $G_i$-good. Let $\pi \in \Gamma$, and let $a \in A$ be such that its 1-type is $\pi$. Let us fix an index $i$ where $1 \leqslant i \leqslant m$. Since $\mathcal{A} \models \forall x \exists y\ \beta_i(x, y)$, there is an element $b \in A$ such that $\mathcal{A}, x/a, y/b \models \beta_i(x, y)$. Moreover, $\mathcal{A}, x/a, y/b \models \alpha(x, y)$. Thus, $\mathcal{A}, x/a, y/b \models \alpha(x, y) \wedge \beta_i(x, y)$. This means that $\pi(x) \wedge \eta(x, y) \wedge \pi'(y) \models \alpha(x, y) \wedge \beta_i(x, y)$, where $\pi'$ and $\eta$ are the 1- and 2-type of $b$ and $(a, b)$, respectively. This also implies that $\pi'$ is realized in $\mathcal{A}$, and by definition, $\pi' \in \Gamma$. Thus, $(\pi, \pi')$ is $\beta_i$-compatible with $\alpha$, which means $(\pi, \pi')$ is an edge in $G_i$. This concludes that $\Gamma$ is a GIS in $\mathcal{G}_\Phi$.

(if) Let $\Gamma$ be a GIS in $\mathcal{G}_\Phi$. We will construct a structure $\mathcal{A} \models \Phi$. First, for each $\pi \in \Gamma$, we fix a countable infinite set $A_\pi$ such that for different $\pi, \pi' \in \Gamma$, the sets $A_\pi$ and $A_{\pi'}$ are disjoint. The universe of $\mathcal{A}$ is the set $A = \bigcup_{\pi \in \Gamma} A_\pi$. Next, we will define the interpretation of each relation symbol by defining the 1-types of each $a \in A$ and the 2-type of each pair $(a, b) \in A \times A$. For 1-types, we set the 1-type of each element in $A_\pi$ to be $\pi$ itself.

The 2-type of every pair $(a, b) \in A \times A$ is defined as follows. We first enumerate all the elements in $A$ as $a_1, a_2, \ldots$. The assignment of the 2-types is done by iterating the following process starting from $i = 1$ to $i \to \infty$. In the $i$-th iteration, we partially assign 2-types of pairs involving $a_i$ as follows.

- Let $\pi$ be the 1-type of $a_i$, i.e., $a_i \in A_\pi$. We pick $m$ elements $a_{i_1}, \ldots, a_{i_m}$ such that for each $1 \leqslant j \leqslant m$:
  - $i_j > i$, and
  - each $a_{i_j} \in A_{\pi_j}$, where $(\pi, \pi_j)$ is an edge in $E_j(\mathcal{G}_\Phi)$.

  Such elements exist due to the fact that $\Gamma$ is a GIS and that each $A_\pi$ is an infinite set.
  By definition, for every $1 \leqslant j \leqslant m$, $(\pi, \pi_j)$ is $\beta_j$-compatible with $\alpha$. Hence, there is 2-type $\eta_j$ such that:

$$\pi(x) \wedge \eta_j(x, y) \wedge \pi_j(y) \quad \models \quad \alpha(x, y) \wedge \beta_j(x, y) \tag{2}$$

  We set the 2-type of $(a_i, a_{i_j})$ to be $\eta_j$, for each $1 \leqslant j \leqslant m$.
- For every $j < i$, whenever the 2-type of $(a_i, a_j)$ is not yet defined, it is defined as follows. Let $\pi' \in \Gamma$ be the 1-type of $a_j$. Then, we pick 2-type $\eta$ such that:

$$\pi(x) \wedge \eta(x, y) \wedge \pi'(y) \quad \models \quad \alpha(x, y) \tag{3}$$

  Such $\eta$ exists, since $\Gamma$ is GIS, and hence, $(\pi, \pi')$ is compatible with $\alpha$. We set the 2-type of $(a_i, a_j)$ to be $\eta$.

As $i \longrightarrow \infty$, the 2-type of every pair is well-defined.

We now show that $\mathcal{A} \models \Phi$. First, $\mathcal{A} \models \forall x \forall y\ \alpha(x, y)$, since for every pair $(a, b) \in A \times A$, we only assign 2-type $\eta$ where either Eq. (2) or Eq. (3) hold, i.e., when $\alpha(x, y)$ is respected. Moreover, the first bullet ensures that for every element $a_i$, the element $a_{i_j}$ is chosen such that $\mathcal{A}, x/a_i, y/a_{i_j} \models \alpha(x, y) \wedge \beta_j(x, y)$. Thus, $\mathcal{A} \models \forall x \exists y\ \beta_j(x, y)$, for every $1 \leqslant j \leqslant m$. This concludes the proof of Lemma 3.1. □

Let $\delta_0$, $\delta_1$ and $\delta_2$ be the constants defined in Theorems 2.2, 2.3 and 2.4, respectively. Combining Lemma 3.1 with the results in Section 2, we obtain the following.

Theorem 3.2. *On the fragment of FO$^2$ without the equality predicate, the following holds.*

- *There is a non-deterministic algorithm for SAT(FO$^2$) with run time $O(2^n)$.*
- *There is a deterministic algorithm for SAT(FO$^2$) with run time $O\big(\delta_0^{(2^n)}\big)$ and $O\big(\delta_1^{(2^n)}\big)$.*
- *Unless SETH fails, every algorithm that decides SAT(FO$^2$) has run time $\Omega\big(\delta_2^{(2^n)}\big)$*

*Here the input formula is in Scott normal form (1) and n is the number of predicates.*

To end this section, we remark that for every graph system $\mathcal{G}$, one can easily construct an FO$^2$ formula $\varphi$ (without the equality predicate and using only unary predicates) such that $\mathcal{G}_\varphi = \mathcal{G}$, and the number of unary predicates used is logarithmic in the number of vertices in $\mathcal{G}$. This way one can view SAT(FO$^2$) as succinct representation of CIS where the complexity is measured in the number of bits needed for naming the vertices in $\mathcal{G}$. Naturally, for some graph systems, the formulas that encode them have length proportional to the number of vertices, even if they use only logarithmic number of unary predicates.

### 3.3 Fragments with lower complexity

In this subsection we present a few fragments of FO$^2$ whose satisfiability problem has complexity lower than NEXP-complete. The results in this subsection parallel those in Subsection 2.5. We start with the following definitions.

*Definition 3.3.* A formula $\Phi$ in Scott normal form (1) is called a *conflict-free*/*uniquely-outgoing* formula, if its graph system $\mathcal{G}_\Phi$ is conflict-free/uniquely-outgoing, respectively.

Note that we can decide in polynomial space if a formula is conflict-free or uniquely-outgoing.

*EXP-complete fragment.* First, we show that restricted on conflict-free formulas, SAT(FO$^2$) drops to EXP-complete. The upper bound follows directly from Theorem 2.9 and the fact that the constructed graph system may have exponentially many vertices.

For The lower bound, we use the fact that alternating polynomial space Turing machines are equivalent to exponential time deterministic Turing machines [6]. Let $M$ be an alternating 1-tape Turing machine that uses $cn$ space, for some $c \geqslant 1$. Let $Q$ be the set of its states and $\Delta$ the tape alphabet. Without loss of generality, we assume that there are exactly two transitions that can be applied on every universal state. Moreover, every configuration always leads to a halting configuration, since we can assume that $M$ has a "counter" that counts the number of steps taken so far and $M$ rejects immediately if the counter reaches $O(2^{cn})$.

For an input word $w = a_1 \cdots a_n \in \{0, 1\}^*$, we construct a formula of the form:

$$\Phi' \quad := \quad \forall x\ \phi_0(x)\ \wedge\ \forall x \exists y\ \phi_1(x, y)\ \wedge\ \forall x \exists y\ \phi_2(x, y)$$

such that $M$ accepts $w$ if and only if $\Phi'$ is satisfiable. Note that since $\Phi'$ does not have the conjunct $\forall\forall$, its graph system will not have any conflicting vertices.

The vocabulary of $\Phi'$ consists of only unary predicates $U_{b,i}$, where $b \in Q \cup \Delta$ and $1 \leqslant i \leqslant cn$. Intuitively, a configuration $b_1 \cdots b_{k-1}(q, b_k)b_{k+1} \cdots b_{cn}$ (i.e., the head is in cell $k$ and in state $q$ and the content of the tape is $b_1 \cdots b_{cn}$) is represented as 1-type whose positive literals are $U_{q,k}(x)$ and $U_{b_i,i}(x)$ for every $1 \leqslant i \leqslant cn$. All other literals are negative.

The formulas $\phi_0(x)$, $\phi_1(x,y)$, $\phi_2(x,y)$ are defined as follows, where $[cn]$ denotes the set $\{1,\ldots,cn\}$.

- $\phi_0(x)$ states that 1-type of $x$ represents a configuration.
  Formally,

$$\Big( \bigvee_{(q,i)\in Q\times[cn]} \Big(U_{q,i} \ \wedge \bigwedge_{(p,j)\neq(q,i) \text{ and } (p,j)\in Q\times[cn]} \neg U_{p,j}(x)\Big)\Big)$$

$$\wedge \ \bigwedge_{i\in[cn]} \bigvee_{b\in\{0,1,\#\}} \Big(U_{b,i}(x) \ \wedge \bigwedge_{b'\neq b \text{ and } b'\in\{0,1,\#\}} \neg U_{b',i}(x)\Big)$$

- $\phi_1(x,y)$ and $\phi_2(x,y)$ state that the configuration represented by the 1-type of $y$ is the next step of the configuration represented by the 1-type of $x$, and we define the "next" step of an accepting configuration to be the initial configuration.
  Formally, they are defined as follows. The formula $\phi_1(x,y)$ handles the next step for the existential configurations:

$$\bigwedge_{(q,i)\in Q\times[cn] \text{ where } q \text{ is existential and } b\in\{0,1,\#\}} \Big(\big(U_{q,i}(x) \ \wedge \ U_{b,i}(x)\big) \ \rightarrow \ \varphi_{\text{next}(q,b,i)}(y)\Big)$$

  where $\varphi_{\text{next}(q,b,i)}(y)$ states that certain unary predicates must hold on $y$ according to the transitions in $M$ when the state is in $q$ and the head is reading symbol $b$. Moreover, if $q = q_{acc}$, then $\varphi_{\text{next}(q,b,i)}(y)$ states that 1-type of $y$ represents the initial configuration.
  To handle universal states, we add the following conjunct in $\phi_1(x,y)$:

$$\bigwedge_{(q,i)\in Q\times[cn] \text{ where } q \text{ is universal and } b\in\{0,1,\#\}} \Big(\big(U_{q,i}(x) \ \wedge \ U_{b,i}(x)\big) \ \rightarrow \ \varphi_{\text{next-1}(q,b,i)}(y)\Big)$$

  and define $\phi_2(x,y)$ as:

$$\bigwedge_{(q,i)\in Q\times[cn] \text{ where } q \text{ is universal and } b\in\{0,1,\#\}} \Big(\big(U_{q,i}(x) \ \wedge \ U_{b,i}(x)\big) \ \rightarrow \ \varphi_{\text{next-2}(q,b,i)}(y)\Big)$$

  Here $\varphi_{\text{next-1}(q,b,i)}(y)$ and $\varphi_{\text{next-2}(q,b,i)}(y)$ denotes the "next" step when the first and second transitions are applied on the universal state $q$.

Note that two formulas $\phi_1(x,y)$ and $\phi_2(x,y)$ are required to define the next step of a configuration with universal state.

It is routine to verify that if $w$ is accepted by $M$, then $\Phi$ has a model that represents its accepting run. On the other hand, if $w$ is rejected by $M$, the run will go to a rejecting configuration, for which there is no "next" step and $\Phi$ does not have any model.

*The two-variable guarded fragment.* In this part we will show that the two-variable guarded fragment formulas (without the equality predicate) fall inside our EXP-complete fragment. Hence, its satisfiability problem can be decided in exponential time. This recovers a special case of Grädel's result [11] that the satisfiability of guarded fragment (with the equality predicate) with a fixed number of variables/arity of the predicates is EXP-complete.

Indeed, the normal form of guarded fragment formulas with two variables are of the form:

$$\forall x\ \gamma(x) \ \wedge \ \bigwedge_{i=1}^{n} \forall x \forall y\ e_i(x,y) \rightarrow \alpha_i(x,y) \ \wedge \ \bigwedge_{i=1}^{m} \forall x \exists y\ f_i(x,y)$$

where $e_i(x,y), f_i(x,y)$ are atomic predicates and $\gamma(x), \alpha_i(x,y)$ are quantifier free. See, e.g., [20, 35]. Due to the guard $e_i(x,y)$, every realized 1-types $\pi, \pi'$ are compatible. Thus, its graph system is conflict-free and our algorithms for SAT$(\mathsf{FO}^2)$ runs in exponential time.

Combining this with our proof above, we obtain the following.

- The satisfiability problem for the formulas of the form $\varphi \wedge \psi$, where $\varphi$ is a guarded fragment formula with two variables and $\psi$ has prefix $\forall \wedge (\forall\exists)^*$ is EXP-complete.
- The satisfiability problem for the formulas with prefix $\forall \wedge \forall\exists$ is PSPACE-complete.

*PSPACE-complete fragment.* Next, we show that on conflict-free formulas with $m = 1$, SAT$(\mathsf{FO}^2)$ drops further to PSPACE-complete. Here $m$ is as in the Scott normal form (1).

For the upper bound, note that by Lemma 2.7, it suffices to check the existence of a cycle in the constructed graph system, which can be done by guessing a vertex and a cycle that contains it. Each vertex is a 1-type, hence requires linear space. Verifying whether there is an edge between two vertices is essentially a Boolean SAT problem, hence, can also be done in linear space.

For the lower bound, the reduction is similar to the one for EXP-hardness, except that $M$ is now a determinstic polynomial space Turing machine. On input word $w = a_1 \cdots a_n \in \{0,1\}^*$, we construct a formula of the form:

$$\Phi'' \quad := \quad \forall x\ \phi_0(x)\ \wedge\ \forall x \exists y\ \phi_1(x,y)$$

where $\phi_0(x)$ and $\phi_1(x,y)$ are as in $\Phi'$. Note that since $M$ is a deterministic Turing machine, i.e., there is no universal state, we only require one formula $\phi_1(x,y)$ to express the next step of a configuration.

*Uniquely-outgoing formulas.* Finally, we show that on uniquely-outgoing formulas, SAT$(\mathsf{FO}^2)$ is PSPACE-complete. The upper bound is obtained by Lemma 2.8, i.e., by checking if there is a vertex $u$ such that $\mathcal{R}(u)$ is a GIS. Note that the (non-deterministic) algorithm that checks whether $\mathcal{R}(u)$ is not a GIS can be employed here. Again, since each vertex is a 1-type that requires linear space, and verifying whether there is an edge between two vertices can done in linear space, overall the algorithm uses only polynomial space.

For the lower bound, note that the constructed $\Phi''$ above yields a uniquely-outgoing graph system. Recall that $\phi_1(x,y)$ states that the 1-type of $y$ represents the next step of the 1-type of $x$. Since $M$ is deterministic, the 1-type of $y$ must be unique. Thus, the constructed graph system of $\Phi''$ is uniquely-outgoing.

We summarize formally the results in this subsection as Theorem 3.4 below.

THEOREM 3.4.

- *SAT$(FO^2)$ is EXP-complete on conflict-free formulas with $m \geqslant 2$.*
- *SAT$(FO^2)$ is PSPACE-complete on conflict-free formulas with $m = 1$.*
- *SAT$(FO^2)$ is PSPACE-complete on uniquely-outgoing formulas.*

## 4 $\mathsf{FO}^2$ WITH THE EQUALITY PREDICATE

In general $\mathsf{FO}^2$ with the equality predicate is more expressive than the fragment without. For example, the formula $\forall x \forall y\ U(x) \wedge U(y) \to x = y$, which semantically states that the predicate $U$ can only hold on at most one element, cannot be expressed even in full first-order logic without using the equality predicate.

In this section we will show how to reduce in linear time a formula with the equality predicate to an equi-satisfiable formula without the equality predicate. We start by observing that every $\mathsf{FO}^2$ formula with the equality predicate in Scott normal form can be further rewritten into the

following form:

$$\Psi \;:=\; \forall x\, \gamma(x) \quad \wedge \quad \forall x \forall y \,\big(x \neq y \rightarrow \alpha(x,y)\big) \quad \wedge \quad \bigwedge_{i=1}^{m} \forall x \exists y\, \beta_i(x,y) \wedge x \neq y \tag{4}$$

where each $\beta_i(x,y)$ is an atomic predicate and $\alpha(x,y)$ does not use the equality predicate.

Note that in Eq. (4) $\beta_i$ is an atomic predicate, whereas in Eq. (1) $\beta_i$ is a quantifier free formula. This is for technical convenience. Without loss of generality, we assume $\Psi$ does not use constant symbols, since they can be represented with unary predicates. We also assume that there is no atom of the form $R(x,x)$ or $R(y,y)$, where $R$ is binary predicate, since they can be treated like unary predicates.

Let $\tau = \{U_1,\ldots,U_n,\beta_1,\ldots,\beta_k\}$ be the vocabulary of $\Psi$, where $k \geqslant m$, and $U_i$ and $\beta_i$ are unary and binary predicates used in $\Psi$, respectively. We will construct a formula $\Psi^*$ equi-satisfiable to $\Psi$. The intuition is as follows. The formula $\Psi^*$ is defined so that if it is satisfiable, then the universe of any of its models can be seen as built of the pairs $(a,b) \in \bigcup_{i=1}^{m} \beta_i^{\mathcal{A}}$, for some structure $\mathcal{A}$ that satisfies $\Psi$. The main challenge is to describe the properties of those pairs without using the equality predicate.

The formula $\Psi^*$ will be defined over vocabulary $\tau^*$ which consists of the following predicates.

- Binary predicates: $\beta_1,\ldots,\beta_k$ (as in $\tau$).
- Unary predicates: $K_s$, $K_t$, $S_1,\ldots,S_n$, $T_1,\ldots,T_n$, $P_1,\ldots,P_k$, $\overline{P}_1,\ldots,\overline{P}_k$, $Z_1,\ldots,Z_{n+\log_2(3m)}$.

For the rest of this section, we denote by $p = n + \log_2 3m$.

The terminology of *king type* and *king element* from [12] will be crucial. So we recall them formally here.

*Definition 4.1.* Suppose $\mathcal{A}$ is a structure over vocabulary $\tau$. We say that a 1-type $\pi$ is a *king type* in $\mathcal{A}$, if there is only one element $a \in A$ with 1-type $\pi$. In this case, the element $a$ is called a *king* element in $\mathcal{A}$.

The notion of *edge representation structure* below is our formalism on how structures over $\tau^*$ correspond to pairs of elements in the predicates $\bigcup_{i=1}^{m} \beta_i$ in structures over $\tau$.

*Definition 4.2.* Let $\mathcal{A}$ be a structure over $\tau$. The *edge representation structure of* $\mathcal{A}$ is a structure $\mathcal{B}$ over vocabulary $\tau^*$ defined as follows.

- The universe $B$ consists of all the pairs $(a,b) \in A \times A$, where $(a,b)$ or $(b,a)$ is in $\beta_i^{\mathcal{A}}$ for some $1 \leqslant i \leqslant m$.
- $(a,b) \in K_s^{\mathcal{B}}$ if and only if $a$ is a king element in $\mathcal{A}$.
- $(a,b) \in K_t^{\mathcal{B}}$ if and only if $b$ is a king element in $\mathcal{A}$.
- For each $1 \leqslant i \leqslant n$,
  - $(a,b) \in S_i^{\mathcal{B}}$ if and only if $a \in U_i^{\mathcal{A}}$, and
  - $(a,b) \in T_i^{\mathcal{B}}$ if and only if $b \in U_i^{\mathcal{A}}$.
- For each $1 \leqslant i \leqslant k$,
  - $(a,b) \in P_i^{\mathcal{B}}$ if and only if $(a,b) \in \beta_i^{\mathcal{A}}$, and
  - $(a,b) \in \overline{P}_i^{\mathcal{B}}$ if and only if $(b,a) \in \beta_i^{\mathcal{A}}$.

As mentioned earlier, since atoms such as $\beta_i(x,x)$ and $\beta_i(y,y)$ are treated as unary predicates, it is implicit that $a \neq b$, for every $(a,b) \in \beta_i^{\mathcal{A}}$, and hence, for every $(a,b) \in B$. Note also that the definition of $\mathcal{B}$ does not contain the interpretation of $\beta_i$'s and $Z_i$'s. Predicates $\beta_i$'s will be used to describe the property of "non-pairs" in $\mathcal{A}$, whereas $Z_i$'s will be used to describe the property of pairs $(a,b)$, where $a$ is non-king and $b$ is king in $\mathcal{A}$.

We need some more terminology. For a structure $\mathcal{A}$ over $\tau$, the terms *source* and *target* of a pair $(a,b) \in A \times A$ refer to the elements in the first and second components in the pair $(a,b)$, respectively, i.e., $a$ and $b$. For the rest of this section, we use $\mathcal{A}$ to denote a structure over $\tau$ and $\mathcal{B}$ a structure over $\tau^*$. As usual, $A$ and $B$ denote the universe of $\mathcal{A}$ and $\mathcal{B}$, respectively.

Let $\pi$ be a 1-type over $\tau^*$. The *S-type* of $\pi$ is defined to be the intersection of $\pi$ and the set $\{S_i(x), \neg S_i(x) \mid 1 \leqslant i \leqslant n\}$. In other words, $S$-types are maximally consistent sets involving only the predicates $S_1, \ldots, S_n$. The notions of *$T$-type* and *$Z$-type* can be defined in similar manner, i.e., involving only the predicates $T_1, \ldots, T_n$ and $Z_1, \ldots, Z_p$, respectively. The *$P$-type* of $\pi$ is the intersection between $\pi$ and the set $\{P_i(x), \overline{P}_i(x), \neg P_i(x), \neg \overline{P}_i(x) \mid 1 \leqslant i \leqslant k\}$. Note that $P$-type involves both the predicates $P_i(x)$'s and $\overline{P}_i(x)$'s.

The $S$-type, $T$-type, $Z$-type and $P$-type of an element $b \in B$ is defined as the $S$-type, $T$-type, $Z$-type and $P$-type of $\pi$, respectively, where $\pi$ is the 1-type of $b$. Intuitively, if an element $b \in B$ represents a pair $(a_1, a_2)$ in $\mathcal{A}$, the $S$-type and $T$-type of $b$ represent the 1-types of the source and target of pair $(a_1, a_2)$, respectively. The $P$-type of $b$ represents the 2-type of pair $(a_1, a_2)$. We will use $Z$-type to denote the "id" of the sources of pairs. Intuitively, if two elements in $B$ have the same $Z$-types, it means that their sources have the same id, which means that they can be regarded as the same element. This will be useful to enforce certain properties on the 2-type of pairs $(a_1, a_2)$, where the source is not a king element, but the target is.

We define some formulas that will be useful later on.

$$\xi_S(x, y) \quad := \quad \bigwedge_{i=1}^{n} S_i(x) \leftrightarrow S_i(y)$$

Intuitively, it states that the $S$-types of $x$ and $y$ are the same. We can define $\xi_T(x, y)$, $\xi_Z(x, y)$ and $\xi_P(x, y)$ in similar manner which state that $x$ and $y$ have the same $T$-type, $Z$-type and $P$-type, respectively.

$$\xi_{\text{rev}}(x, y) \quad := \quad \bigwedge_{i=1}^{k} \big(P_i(x) \leftrightarrow \overline{P}_i(y)\big) \;\wedge\; \big(\overline{P}_i(x) \leftrightarrow P_i(y)\big)$$

$$\xi_{S,T}(x, y) \quad := \quad \bigwedge_{i=1}^{n} S_i(x) \leftrightarrow T_i(y)$$

Intuitively, these two formulas state that $x$ is the "reverse" of $y$ and that the $S$-type of $x$ is the "same" as the $T$-type of $y$, respectively.

Next, the formula $\xi_{\neq}(x, y)$ below states that the sources of $x$ and $y$ represent different elements.

$$\xi_{\neq}(x, y) \quad := \quad \big(K_s(x) \;\wedge\; K_s(y)\big) \;\rightarrow\; \neg \xi_S(x, y)$$

The intuition is as follows. Suppose $b_1, b_2 \in B$ represent the pairs $(a_1, a_1')$ and $(a_2, a_2')$ in $\mathcal{A}$, respectively. Recall that $b_1$ and $b_2$ belong to $K_s$ means that the sources of $b_1$ and $b_2$ are king elements in $\mathcal{A}$. Thus, if $a_1$ and $a_2$ are kings and they have different types, then they are different. Later on, we will see that if at least one of $a_1$ or $a_2$ is not king, we can assume that they are different elements, even when they have the same type.

Finally, we have the formula $\xi_{\text{id}}(x, y)$ defined below.

$$\xi_{\text{id}}(x, y) \quad := \quad \xi_S(x, y) \;\wedge\; \big(\big(\neg K_s(y) \wedge K_t(y)\big) \;\rightarrow\; \xi_Z(x, y)\big)$$

Intuitively, it states that $x$ and $y$ have the same $S$-type and if the source of $y$ is not king, but the target is a king, then $x$ and $y$ have the same $Z$-type.

Now we are ready to define $\Psi^*$:

$$\Psi^* \quad := \quad \bigwedge_{i=1}^{10} \psi_i$$

Below is the definition of each $\psi_i$ together with their intuitive meaning.

$$\begin{array}{lcl}
\psi_1 & := & \forall x \quad \bigvee_{i=1}^{m} P_i(x) \ \vee \ \overline{P}_i(x) \\
\psi_2 & := & \forall x \forall y \quad \xi_S(x,y) \ \rightarrow \ \big(K_s(x) \ \leftrightarrow \ K_s(y)\big) \\
\psi_3 & := & \forall x \forall y \quad \xi_T(x,y) \ \rightarrow \ \big(K_t(x) \ \leftrightarrow \ K_t(y)\big) \\
\psi_4 & := & \forall x \exists y \left( \begin{array}{l} \xi_{S,T}(x,y) \ \wedge \ \xi_{S,T}(y,x) \ \wedge \ \xi_{\mathrm{rev}}(x,y) \\ \wedge \ (K_s(x) \leftrightarrow K_t(y)) \\ \wedge \ (K_t(x) \leftrightarrow K_s(y)) \end{array} \right) \\
\psi_5 & := & \forall x \quad \xi_{S,T}(x,x) \ \rightarrow \ \big(\neg K_s(x) \ \wedge \ \neg K_t(x)\big) \\
\psi_6 & := & \forall x \forall y \left( \begin{array}{l} \xi_S(x,y) \ \wedge \ \xi_T(x,y) \\ \wedge \ \xi_Z(x,y) \ \wedge \ K_t(x) \end{array} \right) \ \rightarrow \ \xi_P(x,y) \\
\psi_7 & := & \forall x \quad K_s(x) \ \rightarrow \ \bigwedge_{i=1}^{p} \neg Z_i(x)
\end{array}$$

The intention of the formulas $\psi_1$–$\psi_7$ is to capture the natural properties of pairs in $\mathcal{A}$ and do not depend on the original formula $\Psi$. Their intuitive meaning is as follows. Suppose $\mathcal{A} \models \Psi$. Formula $\psi_1$ states that we are only interested in pairs $(a_1, a_2) \in \beta_i^{\mathcal{A}}$, where $1 \leqslant i \leqslant m$ and it is essential since $\mathcal{A} \models \forall x \exists y \ \beta_i(x,y) \wedge x \neq y$, for every $1 \leqslant i \leqslant m$. Formula $\psi_2$ states that if the sources of two pairs have the same 1-type, then either both sources are kings or both are not kings. Formula $\psi_3$ states likewise regarding the targets. Formula $\psi_4$ states that every pair must have its "inverse." Formula $\psi_5$ states that if the source and target of a pair have the same 1-type, then both are not king elements. Formula $\psi_6$ states that, for every two pairs, if their sources and targets have the same 1-type and the target is a king and the sources have the same id, then 2-type of the pairs are the same. Finally, formula $\psi_7$ states that all king elements have fixed id.

The rest of the formulas $\psi_8$–$\psi_{10}$ are defined according to the formula $\Psi$. Recall that $\Psi$ is in the form of Eq. (4). Formula $\psi_8$ is defined as follows.

$$\psi_8 \quad := \quad \forall x \ \gamma_1(x)$$

where $\gamma_1(x)$ is the formula obtained from $\gamma(x)$ by replacing every atom $U_i(x)$ with $S_i(x)$, for every $1 \leqslant i \leqslant n$. The intention of $\psi_8$ is to represent $\forall x \gamma(x)$.

Formula $\psi_9$ is defined as follows.

$$\psi_9 \quad := \quad \forall x \ \alpha_1(x) \quad \wedge \quad \forall x \forall y \ \xi_{\neq}(x,y) \rightarrow \alpha_2(x,y)$$

where $\alpha_1(x)$ and $\alpha_2(x,y)$ are as follows.

- $\alpha_1(x)$ is obtained from $\alpha(x,y)$ by replacing every atom $U_i(x)$ with $S_i(x)$, every $U_i(y)$ with $T_i(x)$, every $\beta_i(x,y)$ with $P_i(x)$ and every $\beta_i(y,x)$ with $\overline{P}_i(x)$.
- $\alpha_2(x,y)$ is the formula obtained from $\alpha(x,y)$ by replacing every atom $U_i(x)$ with $S_i(x)$ and every $U_i(y)$ with $S_i(y)$. The binary literals stay the same.

The intention is to represent the part $\forall x \forall y (x \neq y \rightarrow \alpha(x,y))$.

Finally, formula $\psi_{10}$ is defined as follows.

$$\psi_{10} \quad := \quad \bigwedge_{i=1}^{m} \forall x \exists y \quad \xi_{\mathrm{id}}(x,y) \ \wedge \ P_i(y)$$

It is intended to represent the part $\bigwedge_{i=1}^{m} \forall x \exists y \ \beta_i(x,y) \wedge x \neq y$.

Remark 2. It is obvious that $\Psi^*$ can be constructed in linear time in the length of $\Psi$. Note that the number of unary and binary predicates in $\Psi^*$ are $3n + 2k + 2 + \log_2 m$ and $k$, respectively, where $n$ and $k$ are the number of unary and binary predicates in $\Psi$.

Moreover, the conjunction of $\psi_1$–$\psi_9$, except $\psi_4$, can be combined into one $\forall\forall$ conjunct. Thus, the constructed $\Psi^*$ is in the form: $\forall x \forall y\ \alpha'(x,y)\ \wedge\ \bigwedge_{i=1}^{m+1} \forall x \exists y\ \beta'_i(x,y)$, which is Scott normal form. Note that the number of $\forall\exists$ conjuncts increases by 1 due to $\psi_4$.

The rest of this section is devoted to the proof that $\Psi$ and $\Psi^*$ are indeed equi-satisfiable, stated as Lemmas 4.3 and 4.4 below.

Lemma 4.3. *If $\Psi$ is satisfiable, then $\Psi^*$ is satisfiable.*

Proof. Let $\mathcal{A} \models \Psi$. By the ESM property of $\mathsf{FO}^2$ [12, Theorem 4.3], we assume that $|A| \leqslant 3m2^n$. Let $\mathcal{B}$ be the edge representation structure of $\mathcal{A}$. We first need to define the interpretation of each predicate $\beta_i$ and $Z_i$ in $\mathcal{B}$. For $\beta_i$, it suffices to define the 2-types of every pair $((a_1,b_1),(a_2,b_2))$ in $\mathcal{B}$. There are three cases.

- Case 1: $a_1 = a_2 = a$ and $a$ is a king element in $\mathcal{A}$.
  In this case, both $((a,b_1),(a,b_2))$ and $((a,b_2),(a,b_1))$ are not in $\beta_i^{\mathcal{B}}$, for every $1 \leqslant i \leqslant k$. In other words, the 2-type of $((a_1,b_1),(a_2,b_2))$ contains the literals $\neg\beta_i(x,y)$ and $\neg\beta_i(y,x)$, for every $1 \leqslant i \leqslant k$.
- Case 2: $a_1 = a_2 = a$ and $a$ is not a king element in $\mathcal{A}$.
  In this case, let $a' \in A$ be an element such that $a' \neq a$, but $a'$ has the same 1-type as $a$. We define the 2-type of $((a,b_1),(a,b_2))$ in $\mathcal{B}$ as the 2-type of $(a,a')$ in $\mathcal{A}$.
  Note that since 2-type of $(a,a')$ uniquely determine its "inverse" $(a',a)$, it is is implicit that the 2-type of $((a,b_2),(a,b_1))$ in $\mathcal{B}$ is the 2-type of $(a',a)$ in $\mathcal{A}$.
- Case 3: $a_1 \neq a_2$.
  In this case, we define the 2-type of $((a_1,b_1),(a_2,b_2))$ in $\mathcal{B}$ as the 2-type of $(a_1,a_2)$ in $\mathcal{A}$.

Now we define the interpretation of each predicate $Z_i$ in $\mathcal{B}$.

- For each king element $a \in A$, we define $(a,b)$ *not* to be in $Z_i^{\mathcal{B}}$, for every $Z_i$.
- For each non-king element $a \in A$, we choose a subset $Q_a \subseteq \{Z_1,\ldots,Z_p\}$ such that for different $a,a' \in A$, the sets $Q_a$ and $Q_{a'}$ are different. Note that since $|A| \leqslant 3m2^n$, such sets $Q_a$'s exist.
  Then, for every $(a,b) \in B$, we define $(a,b)$ to be in $Z_i^{\mathcal{B}}$ if and only if $Z_i \in Q_a$.

Now, we will show that $\mathcal{B} \models \Psi^*$. It is routine to verify that $\mathcal{B}$ satisfies the formulas $\psi_1$–$\psi_7$. So we will only show that $\mathcal{B} \models \psi_8 \wedge \psi_9 \wedge \psi_{10}$. We first show that for every $(a,b),(a_1,b_1),(a_2,b_2) \in B$, the following holds.

(a) $\mathcal{B}, x/(a,b) \models \gamma_1(x)$.
(b) $\mathcal{B}, x/(a,b) \models \alpha_1(x)$.
(c) $\mathcal{B}, x/(a_1,b_1), y/(a_2,b_2) \models \xi_{\neq}(x,y) \to \alpha_2(x,y)$.

Recall that for $(a,b) \in B$, we have $a \neq b$. To prove (a), note that by definition, $\gamma_1(x)$ is exactly the same formula as $\gamma(x)$, except that each $U_i(x)$ is replaced by $S_i(x)$. Now since the $S$-type of $(a,b)$ is exactly the 1-type of $a$. and that $\mathcal{A}, x/a \models \gamma(x)$, we have $\mathcal{B}, x/(a,b) \models \gamma_1(x)$. The proof of (b) is similar.

To prove (c), suppose $\mathcal{B}, x/(a_1,b_1), y/(a_2,b_2) \models \xi_{\neq}(x,y)$. There are two cases.

- Case 1: $a_1 \neq a_2$.
  By construction of $\mathcal{B}$, the 2-type of $((a_1,b_1),(a_2,b_2))$ in $\mathcal{B}$ is exactly the 2-type of $(a_1,a_2)$ in $\mathcal{A}$. Moreover, the $S$-types of $(a_1,b_1)$ and $(a_2,b_2)$ in $\mathcal{B}$ are exactly the 1-type of $a_1$ and $a_2$ in $\mathcal{A}$, respectively. Since $\alpha_2(x,y)$ is exactly the same formula as $\alpha(x,y)$, except that each $U_i(x)$ and $U_i(y)$ are replaced by $S_i(x)$ and $S_i(y)$, respectively, it follows that $\mathcal{B}, x/(a_1, b_1), y/(a_2, b_2) \models \alpha_2(x, y)$.

- Case 2: $a_1 = a_2 = a$ and it is not a king element.
  By construction of $\mathcal{B}$, the 2-type of $((a, b_1), (a, b_2))$ in $\mathcal{B}$ is the 2-type of $(a, a')$ in $\mathcal{A}$, for some $a' \neq a$ with the same 1-type as $a$. With the same reasoning as above, it follows that $\mathcal{B}, x/(a, b_1), y/(a, b_2) \models \alpha_2(x, y)$.

(a)–(c) above immediately implies that $\mathcal{B} \models \psi_8 \wedge \psi_9$. To show that $\mathcal{B} \models \psi_{10}$, let $(a, b) \in B$ and $1 \leqslant i \leqslant m$. Since $\mathcal{A} \models \Psi$, there is $b'$ such that $\mathcal{A}, x/a, y/b' \models \beta_i(x, y) \wedge x \neq y$. By the construction of $\mathcal{B}$, we have:

$$\mathcal{B}, x/(a, b), y/(a, b') \models \xi_S(x, y) \wedge P_i(y) \qquad \text{and} \qquad \mathcal{B}, x/(a, b), y/(a, b') \models \xi_Z(x, y)$$

Thus, $\mathcal{B}, x/(a, b), y/(a, b') \models \xi_{\text{id}}(x, y) \wedge P_i(y)$. This completes our proof of Lemma 4.3. □

Lemma 4.4. *If $\Psi^*$ is satisfiable, then $\Psi$ is satisfiable.*

Proof. Suppose $\mathcal{B} \models \Psi^*$. We say that an $S$-type $\pi$ is *realized* in $\mathcal{B}$, if there is an element $b \in B$ with $S$-type $\pi$. Moreover, it is a king $S$-type in $\mathcal{B}$, if there is an element $b \in B$ with $S$-type $\pi$ and $b \in K_s^{\mathcal{B}}$. Note that since $\mathcal{B} \models \psi_2$, if there is such an element $b$, every element with the same $S$-type as $b$ belongs to the predicate $K_s^{\mathcal{B}}$.

In the following, since $S$-types and $T$-types correspond to 1-types over vocabulary $\tau$, when there is no confusion, abusing the notation, we will often use the terms $S$-type/$T$-type interchangably with 1-type over $\tau$. For example, suppose $\pi$ is an $S$-type. We say that "*the 1-type of an element $a$ in $\mathcal{A}$ is $\pi$*" when we mean that its 1-type contains atom $U_i(x)$ if and only if $S_i(x)$ is in $\pi$, for every $1 \leqslant i \leqslant n$. Likewise, for $T$-types.

We construct a structure $\mathcal{A}$ over vocabulary $\tau$ as follows. For every realized $S$-type $\pi$ in $\mathcal{B}$, we pick a set $A_\pi$ such that the following holds.

- If $\pi$ is a king $S$-type, then $A_\pi$ is a singleton.
- Otherwise, $A_\pi$ is an infinite countable set.

Note that $A_\pi$ is well-defined, since $\mathcal{B} \models \psi_2$. Moreover, we pick two disjoint sets $A_\pi$ and $A_{\pi'}$, for every two different $S$-types $\pi$ and $\pi'$. The universe of $\mathcal{A}$ is defined as the union of all $A_\pi$, where $\pi$ ranges over the $S$-types realized in $\mathcal{B}$.

We now define the interpretation of the predicate $U_i$ and $\beta_i$ in $\mathcal{A}$. To define the interpretation of each $U_i$ in $\mathcal{A}$, it suffices to define the 1-type of each element. For every $\pi$, the 1-type of each element in $A_\pi$ is $\pi$. Note that since $\mathcal{B} \models \forall x\ \gamma_1(x)$, it follows immediately that $\mathcal{A} \models \forall x\ \gamma(x)$.

Next, we define the interpretation of each $\beta_i$ in $\mathcal{A}$. In the following we will refer to the element in a singleton $A_\pi$ as a king element, whereas elements in infinite set $A_\pi$ as non-king elements. Let $\mathcal{K}$ be the set of all king elements, and let $\mathcal{P}$ be the set of all non-king elements. We will define the 2-type of every pair $(a_1, a_2) \in A \times A$, where $a_1 \neq a_2$ according to the three steps outlined below.

(*Step 1*) The goal of this step is to assign 2-types involving elements in $\mathcal{K}$ so that $\mathcal{A}, x/a \models \bigwedge_{i=1}^{m} \exists y\ \beta_i(x, y) \wedge x \neq y$, for every element $a \in \mathcal{K}$.

Let $a \in \mathcal{K}$. Let $\pi$ be the $S$-type where $a \in A_\pi$. Since $\mathcal{B} \models \psi_{10}$, for every $1 \leqslant i \leqslant m$, there is an element $b_i \in B$ such that $\mathcal{B}, x/a, y/b_i \models \xi_{\text{id}}(x, y)$ and $b_i \in P_i^{\mathcal{B}}$. Since $\mathcal{B} \models \psi_4$, for every such $b_i$, there is $b_i' \in B$ whose $S$-type is the same as the $T$-type of $b_i$. Note also that if the $T$-type of $b_i$ is king $T$-type, then since $\mathcal{B} \models \psi_3$, the $S$-type of $b_i'$ is a king $S$-type. Let $a_i$ be an element in $\mathcal{A}$ whose 1-type is the $S$-type of $b_i'$.

Let $W(a)$ denote the set $\{a_1, \ldots, a_m\}$. For each $a_i$, we define the 2-type of $(a, a_i)$ as follows. For every $1 \leqslant j \leqslant k$, the following holds.

- If $b_i \in P_j^{\mathcal{B}}$, it contains $\beta_j(x, y)$, else it contains $\neg\beta_j(x, y)$;

- If $b_i \in \overline{P}_j^{\mathcal{B}}$, it contains $\beta_j(y,x)$, else it contains $\neg\beta_j(y,x)$.

We do this process for every element $a \in \mathcal{K}$. Note that if $a_i \in W(a)$ is also a king element, the 2-type defined on the pair $(a, a_i)$ is well-defined. This is because $\mathcal{B} \models \psi_4 \wedge \psi_5 \wedge \psi_6 \wedge \psi_7$, thus, there is only one 2-type between $a$ and $a_i$. Note also that since $\mathcal{B} \models \psi_8$, and in particular, $\mathcal{B} \models \forall x \alpha_1(x)$, we have that $\mathcal{A}, x/a, y/a_i \models x \neq y \rightarrow \alpha(x,y)$.

For convenience, we also assume the sets $W(a)$'s are defined so that for different $a, a' \in \mathcal{K}$, the sets $W(a) \cap \mathcal{P}$ and $W(a') \cap \mathcal{P}$ are disjoint. Note that this assumption can always hold, since for every non-king $S$-type $\pi$, we define $A_\pi$ to be infinite and there are only finitely many king elements.

(*Step 2*) The goal of this step is to assign 2-types so that $\mathcal{A}, x/a \models \bigwedge_{i=1}^{m} \exists y\ \beta_i(x,y) \wedge x \neq y$, for every element $a \in \mathcal{P}$.

Let $\mathcal{W}$ denote the set $\bigcup_{a \in \mathcal{K}} W(a) \cap \mathcal{P}$. We first achieve the goal for the elements from the set $\mathcal{W}$. Note that the set $\mathcal{W}$ contains all the elements $a$ such that there is exactly one $a' \in \mathcal{K}$ where the 2-type of $(a, a')$ in $\mathcal{A}$ is already defined in Step 1.

Let $a \in \mathcal{W}$ and let $a' \in \mathcal{K}$ be the element such that 2-type of $(a, a')$ is already defined. Let $b \in B$ be an element such that its $S$-type, $T$-type and $P$-type are the 1-type of $a$, 1-type of $a'$ and 2-type of $(a, a')$ in $\mathcal{A}$, respectively. Since $\mathcal{B} \models \psi_{10}$, for every $1 \leqslant i \leqslant m$, there is $b_i \in B$ such that: $\mathcal{B}, x/b, y/b_i \models \xi_{\text{id}}(x,y)$ and $b_i \in P_i^{\mathcal{B}}$. Now, for each $b_i$, let $a_i \in A$ be an element with 1-type the same as the $T$-type of $b_i$. If $a_i$ is not a king element, we can assume that $a_i$ is an element in $\mathcal{P}$ such that 2-types involving $a_i$ is not defined yet. Recall that for non-king type $\pi$, $A_\pi$ is infinite. Thus, such $a_i$ always exists. Then, the 2-type of $(a, a_i)$ can then be defined in similar manner as in Step 1. That it is well defined is established by similar reasoning.

For every element in $a \in \mathcal{P} \setminus \mathcal{W}$, 1-types involving $a$ can also be defined similarly, except that we require from $b \in B$ that its $S$-type is the same as 1-type of $a$. Note that after this step, $\mathcal{A} \models \bigwedge_{i=1}^{m} \forall x \exists y\ \beta_i(x,y) \wedge x \neq y$. Moreover, since $\mathcal{B} \models \forall x\ \alpha_1(x)$, for every pair $(a, a') \in A \times A$, $\mathcal{A}, x/a, y/a' \models x \neq y \rightarrow \alpha(x,y)$.

(*Step 3*) The last step is to define the 2-type of the remaining pairs $(a, a') \in A \times A$. Let $(a, a') \in A \times A$ be a pair whose 2-type yet to be defined. Let $\pi$ and $\pi'$ be the 1-types of $a$ and $a'$, respectively. Then, the 2-type of $(a, a')$ is defined as $\eta(x,y)$ where:

$$\pi(x) \wedge \eta(x,y) \wedge \pi'(y) \wedge x \neq y \quad \models \quad \alpha(x,y)$$

Since $\mathcal{B} \models \forall x \forall y\ \xi_{\neq}(x,y) \rightarrow \alpha_2(x,y)$, such $\eta(x,y)$ always exists for every $\pi$ and $\pi'$. This concludes the proof of Lemma 4.4. □

## 5 EXPERIMENTS

We perform some small experiments comparing our approach with the one in [19, 24]. We implement our $\mathsf{FO}^2$ solver that works as follows. The input formula is in Scott normal form (1) or (4), i.e., without or with the equality predicate, respectively. When it is in form (1), it constructs the graph system first and then runs one of the CIS algorithms. When it is in form (4), it first performs the reduction in Section 4 before proceeding as in the case of (1).

Our solver is implemented in C++ with gcc 9.3.0 (Ubuntu 9.3.0-10ubuntu2) and perform the experiments on i7-6700 CPU @ 3.40GHz with 4 Cores and 8 CPUs and 8GB memory. The OS is Ubuntu 20.04 LTS. We use Z3 [2] version 4.8.7 for solving Boolean SAT when constructing the graph systems.

Below are some snapshots of our experiments where our solvers are compared with the one developed in [19, 24] and Z3 solver. For discussion on how Z3 can be used directly for $\mathsf{SAT}(\mathsf{FO}^2)$, see [19]. The time taken by our solver includes the construction of the graph systems. TO stands for “time out” (24 hours), OM for “out of memory,” UN for “unknown” (when Z3 gives up analyzing and declares “unknown”), and ER for “Exceeds maximal Recursion depth.” We record – when an experiment is not performed since the result should be apparent. For more experiments and their detailed commentary, we refer the reader to the arXiv version of this paper [26].

*Experiment A (without the equality predicate).* The formula is taken from [19] where it is called *2col.* The vocabulary is $\{U_1, \ldots, U_n, V, E\}$, where $U_i, V$ are unary and $E$ binary. Formula $\mathcal{E}_n^A$ states that:

- The sets $U_1, \ldots, U_n$ are pairwise disjoint, and each of them is not empty.
- For every $1 \leqslant i \leqslant n$, for every $j \neq i+1$, there is no $(a,b) \in E$ such that $a \in U_i$ and $b \in U_j$. (Here $n+1$ is defined to be 1.)
- The set $U_1$ is a subset of $V$.
- For every $a$, there is $b$ such that $(a,b) \in E$ and $b \in U_1 \cup \cdots \cup U_n$.
- For every $(a,b) \in E$, $a \in V$ if and only if $b \notin V$.

$\mathcal{E}_n^A$ is satisfiable if and only if $n$ is even. If satisfiable, the smallest model has cardinality $n$. The results are shown in the following table where $n$ is the number of unary predicates.

| | $n$ | Run time on $\mathcal{E}_n^A$ | | | |
|---|---|---|---|---|---|
| | | [19, 24] | Alg-A | Alg-B | Z3 |
| sat. | 6 | 5.74s | 0.1s | 0.1s | UN |
| | 14 | 10h 37m 4s | 0.4s | 0.4s | OM |
| | 30 | OM | 2.5s | 2.5s | OM |
| | 200 | – | 37m 29s | 37m 25s | – |
| | 500 | – | TO | TO | – |
| unsat. | 3 | TO | 0.04s | 0.04s | 0.02s |
| | 13 | – | 0.32s | 0.32s | 2m 9s |
| | 31 | – | 2.55s | 2.55s | OM |
| | 201 | – | 36m 55s | 36m 50s | – |

*Experiment B (without the equality predicate).* The vocabulary is $\{U_1, \ldots, U_{2n}\}$, where each $U_i$ is unary predicate. The formula $\mathcal{E}_n^B$ is defined so that its graph system $\mathcal{G} = (G_0, G_1)$ has $2^{2n}$ vertices and the conflict graph $G_0$ is the Moon-Moser graph [28]. That is, $G_0$ has around $\delta_0^{(2^{2n})}$ number of maximal independent sets, the maximum number possible. Moreover, $G_1$ is defined such that $\mathcal{G}$ has only one GIS. This formula is satisfiable for every $n \geqslant 2$ and the smallest model has cardinality $\lfloor 2^{2n}/3 \rfloor$.

In this instance Alg-A is sensitive towards the "ordering" of the maximal independent set and the unique GIS is contained inside one of the "last" few maximal independent sets. Thus, Alg-A performs rather poorly. The results are shown in the following table where $2n$ is the number of unary predicates.

| $2n$ | Run time on $\mathcal{E}_n^B$ | | | |
|---|---|---|---|---|
| | [19, 24] | Alg-A | Alg-B | Z3 |
| 4 | 3.84s | 0.09s | 0.09s | 0.02s |
| 6 | TO | TO | 0.49s | UN |
| 12 | – | – | 18m 44s | UN |
| 14 | – | – | 4h 56m 3s | UN |
| 16 | – | – | TO | UN |

*Experiment C (with the equality predicate).* The vocabulary is $\{U_1, \ldots, U_n\}$. Formula $\mathcal{E}_n^C$ states that:

- For every element, there is another element whose 1-type is the successor of the 1-type of the former.
- Every 1-type is realizable only on one element.

This formula uses equality predicate and it is satisfiable only by a model with size $2^n$. The results are shown in the following table where $n$ is the number of unary predicates.

| $n$ | Run time on $\mathcal{E}_n^C$ | | | |
|---|---|---|---|---|
| | [19, 24] | Alg-A | Alg-B | Z3 |
| 3 | 3.82s | 0.73s | 0.73s | 0.03s |
| 4 | TO | 3.04s | 3.05s | 0.1s |
| 5 | – | 13s | 13.07s | UN |
| 7 | – | 8m 21s | 8m 20s | UN |
| 8 | – | 2h 29m 17s | 2h 30m 49s | UN |

*Experiment D (random formulas without the equality predicate).* In this experiment we use random $\mathsf{FO}^2$ formulas, which are obtained by first generating random graph systems $\mathcal{G} = (G_0, G_1)$, and then contructing the corresponding $\mathsf{FO}^2$ formulas. Both $G_0$ and $G_1$ are generated independently using the Erdös-Rényi model where the probability of an edge is $1/2$.

The constructed formula is of the form: $\forall x \forall y\ \alpha(x,y)\ \wedge\ \forall x \exists y\ \beta(x,y)$, where $\alpha(x,y)$ and $\beta(x,y)$ are both in CNF. This is to avoid "explicit" listing of the edges inside $\alpha(x,y)$ and $\beta(x,y)$. This way, our solver does not get the edges in $\mathcal{G}$ for free, since it still needs to solve Boolean SAT to obtain them. The results are shown in the following table where $n$ is the number of unary predicates. The number of vertices in $\mathcal{G}$ is always $2^n$.

| $n$ | sat./unsat. | Run time on $\mathcal{E}_n^D$ | | | |
|---|---|---|---|---|---|
| | | [19, 24] | Alg-A | Alg-B | Z3 |
| 2 | unsat. | 3.51s | 0.03s | 0.03s | 0.01s |
| | sat. | 2.19s | 0.03s | 0.03s | UN |
| 5 | sat. | 38.00s | 0.36s | 0.35s | UN |
| | sat. | 50.54s | 0.36s | 0.36s | UN |
| 7 | sat. | ER | 18.38s | 18.43s | UN |
| | sat. | ER | 20.16s | 20.09s | 0.51s |
| 8 | sat. | ER | 6m 53s | 6m 53s | 15m 52s |
| | sat. | ER | 7m 47s | 7m 48s | 5.12s |

## 6 CONCLUDING REMARKS

In this paper we present a novel graph-theoretic approach to $\mathsf{SAT}(\mathsf{FO}^2)$, which yield more efficient algorithms and new intuitions on $\mathsf{FO}^2$. Our approach also gives us a few fragments of $\mathsf{SAT}(\mathsf{FO}^2)$ that come with lower complexity, as well as fresh ideas on how to design interesting benchmarks. In the future we plan to design more benchmarks.

While experimental results seem to validate our algorithms, we also note that our current implementation is a rather naïve one, where the whole graph system is explicitly constructed before running our CIS algorithms. This construction is the main bottleneck of our current implementation, since the CIS algorithms actually run pretty fast. It should be possible that the graph system is constructed only as needed, similar to the tableaux style algorithms employed for other logics [37]. We leave this for future work.